\documentclass[aps,pre,preprint,groupedaddress]{revtex4-1}

\usepackage{url}

\usepackage{comment}
\usepackage{amsmath, amssymb} 
\usepackage{bm}
\usepackage{mathtools}
\usepackage[pdftex]{graphicx, color}

\usepackage{array}

\begin{document}

\title{Factor analysis for a mixture of continuous and binary random variables}

\author{Takashi Arai}
\email[]{takashi-arai@sci.kj.yamagata-u.ac.jp}

\affiliation{Faculty of Science, Yamagata University, Yamagata 990-8560, Japan}

\begin{abstract}
    We propose a multivariate probability distribution that models a linear correlation between binary and continuous variables.
    The proposed distribution is a natural extension of the previously developed multivariate binary distribution.
    As an application of the proposed distribution, we develop a factor analysis for a mixture of continuous and binary variables.
    We also discuss improper solutions associated with factor analysis.
    As a prescription to avoid improper solutions, we propose a constraint that each row vector of factor loading matrix has the same norm.
    We numerically validated the proposed factor analysis and norm constraint prescription by analyzing real datasets.
\end{abstract}

\maketitle

\section{Introduction}
In data analysis, binary random variables appear as often as continuous random variables, for example, in demographic attributes such as gender, in two-point scales of questionnaire responses such as yes/no answers, and in encoding mutations of DNA and amino acid sequences.
Furthermore, real data often contain a mixture of binary and continuous variables, thus one has to deal with binary and continuous variables together.

Binary variables are often encoded into dummy variables that take discrete values in $\{0,1\}$ or $\{-1,1\}$.
Then, the dummy variables are often treated as continuous numeric values in practice, ignoring their discreteness.
Such treatment of binary variables is called methods of quantification.
The quantification allows us to apply many statistical learning methods, such as regression analysis, principal component analysis (PCA), factor analysis, and support vector machines~\cite{Bishop2006, Murphy2012}, to data containing binary variables. 
However, these statistical methods were originally developed to deal with continuous explanatory variables.
Therefore, there is no theoretical justification for such a quantification.
The reason that the quantification is used is a practical compromise due to the lack of statistical methods for properly handling binary variables.
The method of quantification has been used simply because it is convenient in that it is computationally feasible and outputs some results.
Hence, there is a criticism that intrinsically discrete variables are unreflectively treated as continuous numeric variables, and thus, it is debatable whether the results of quantification are meaningful~\cite{stack}.

This paper proposes a multivariate probability distribution that models linear correlation between continuous and binary variables.
Recently, we have succeeded in constructing a multivariate probability distribution for binary variables using Grassmann numbers, anticommuting numbers~\cite{Arai2021}.
We shall refer to the distribution in the previous study as the Grassmann distribution.
The Grassmann distribution has nice properties similar to the multivariate normal distribution and has a computational advantage over the Ising model, a conventional multivariate Bernoulli distribution, in that there is no need to sum over all possible states explicitly when computing the partition function.
The proposed distribution in this paper is a natural extension of the Grassmann distribution.
As an application of the proposed distribution, we construct factor analysis for a mixture of continuous and binary observed variables.
We see that the proposed factor analysis has a computational advantage over existing binary Factor Analysis~\cite{Tipping1998} or exponential family PCA~\cite{Collins2001, Mohamed2008} in estimating model parameters.

This paper is organized as follows.
In Sec.~\ref{sec:result}, we summarize the properties of the proposed distribution.
By fitting the model to real data, we see that the model successfully reproduces a linear correlation between continuous and binary variables.
In Sec.~\ref{sec:fa}, we develop factor analysis for binary random variables as an application of our probability distribution.
We also propose a prescription for avoiding improper solutions of model parameters associated with maximum likelihood estimation of conventional factor analysis. 
The validity of the proposed factor analysis is demonstrated by analyzing real datasets. 
A biplot visualization and its interpretation are given.
Sec.~\ref{sec:conclusion} is devoted to conclusions.

\section{Statement of the results \label{sec:result}}
We denote $p$-dimensional continuous variables and $q$-dimensional binary variables by columns vectors $\mathbf{x}$ and $\mathbf{y}$, respectively.
The binary variables are encoded as dummy variables taking the value $0$ or $1$.
That is, the vector $\mathbf{y}$ is a bit vector with each element taking the value $0$ or $1$.
Model parameters of our distribution consist of mean and covariance parameters of a multivariate normal distribution $(\bm{\mu}, \Sigma)$, a $q \times q$ matrix of the Grassmann distribution $\Lambda$~\cite{Arai2021}, and a $q \times p$ matrix $G$ representing interaction between continuous and binary variables.
The matrix $\Lambda - I$ must be a $P_0$ matrix~\cite{Tsatsomeros2002}, where $I$ is an identity matrix.
Each element of the matrix $G$, $[G]_{sj}, \; (s=1,2,\dots, q, \; \text{and} \; \; j=1,2,\dots,p)$, is also represented by a $p$-dimensional column vector $\mathbf{g}_s$ as
\begin{align}
    G \equiv \begin{bmatrix} \;\;\; \mathbf{g}_1^T \;\;\;\;\; \\  \mathbf{g}_2^T \\ \vdots \\ \mathbf{g}_q^T \end{bmatrix}
    , \hspace{0.5cm} 
    [G]_{sj} = [\mathbf{g}_{s}^T]_j,
\end{align}
where $T$ stands for matrix transposition.
We denote the set of whole indices of continuous and binary variables as $I\equiv \{1,2,\dots,p\}$ and $R \equiv \{1,2,\dots, q\}$, respectively.
An index label for binary variables is divided into two parts with subscripts $1$ and $0$, for variables that take the value $1$ and $0$, respectively.
For example, an index label for binary variables $R$ is divided into a subset $R_1 \subseteq R$ and its set difference $R_0 = R \setminus R_1$.
We denote a $q$-dimensional constant vector $\bm{1}_{R_1}$ with each element taking the value $0$ or $1$,
\begin{align}
    [\bm{1}_{R_1}]_s \equiv
    \begin{cases} 1, \hspace{0.5cm} \text{if} \; \; s \in R_1 \\ 
        0, \hspace{0.5cm} \text{if} \; \; s \in R_0
    \end{cases}, \; \; (s=1, 2, \dots, q).
\end{align}
Then, the proposed jont distribution is expressed as
\begin{align}
    p(\mathbf{x}, \mathbf{y}=\bm{1}_{R_1}) 
    =&  \pi_{R_1}(\Sigma) \, 
    \mathcal{N}(\mathbf{x} \mid \bm{\mu} + \Sigma G^T \mathbf{y}, \Sigma), \notag \\
    \equiv & \pi_{R_1}(\Sigma) 
    \frac{1}{(2\pi)^{p/2} \det \Sigma^{1/2}} \, e^{-\frac{1}{2}(\mathbf{x}- \bm{\mu} - \Sigma G^T \mathbf{y})^T \Sigma^{-1}(\mathbf{x}- \bm{\mu} - \Sigma G^T \mathbf{y})},
    \label{eq:joint} \\
    \pi_{R_1}(\Sigma) \equiv& \frac{\det (\Lambda_{R_0 R_0} -I) e^{\frac{1}{2} \bm{1}_{R_1}^T G \Sigma G^T \bm{1}_{R_1}}}{\sum_{R_1' \subseteq R} \det (\Lambda_{R_0' R_0'} -I) e^{\frac{1}{2} \bm{1}_{R_1'}^T G \Sigma G^T \bm{1}_{R_1'}}},
\end{align}
where $\Lambda_{R_0 R_0}$ is a submatrix of $\Lambda$, and summation $\sum_{R_1' \subseteq R}$ runs over all possible states of binary variables.
The partition function, the normalization constant, of this distribution is not given analytically, and thus, one has to sum over all possible states of the binary variables to calculate the partition function.
As we will see below, the coefficient $\pi_{R_1}(\Sigma)$ corresponds to mixing weight of a mixture of Gaussian distributions with equal covariance.
That is, the above joint distribution corresponds to one normal distribution out of $2^q$ mixture of normal distributions.

To express the marginal and conditional distributions, we first define the notation of index.
We denote the index label of a subset of whole indices as $J \subseteq I$.
Then, the subvector comprising the subset of indices $J$ is represented as $\mathbf{x}_J$.
We divide the sets of whole indices of continuous and binary variables into three subset parts; $I=(J, L, K)$ and $R=(S, U, T)$, where the index labels $L$ and $U$ are introduced to handle missing values.
The number of elements in these sets of indices is represented by $p_J$, $p_L$, $p_K$ and $q_S$, $q_U$, $q_T$, these of course satisfy $p_J + p_L + p_K = p$ and $q_S + q_U + q_T = q$.
Then, the vectors $\mathbf{x}$ and $\mathbf{y}$ can be partitioned into subvectors as $\mathbf{x}=\mathbf{x}_I = (\mathbf{x}_J, \mathbf{x}_L, \mathbf{x}_K)$ and $\mathbf{y}=\mathbf{y}_R=(\mathbf{y}_S, \mathbf{y}_U, \mathbf{y}_T)$, respectively.
Again, an index label for binary variables is further divided into two parts with subscripts $1$ and $0$, for variables that take the value $1$ and $0$, respectively.
For example, an index label for binary variables $S \subseteq R$ is divided into a subset $S_1 \subseteq S$ and its set difference $S_0 = S \setminus S_1$, where the subvectors $\mathbf{y}_{S_1}$ and $\mathbf{y}_{S_0}$ take the values as $\mathbf{y}_{S_1}=\bm{1}$ and $\mathbf{y}_{S_0}=\bm{0}$, respectively.
The union of the index label $J$ and $K$ is denoted as $J + K \equiv J \cup K$.
Using the index notation described above, the marginal distribution is expressed as
\begin{align}
    p(\mathbf{x}_K, \mathbf{y}_T) =& \int d\mathbf{x}_{J+L} \sum_{S_1 +  U_1 \subseteq R \setminus T} p(\mathbf{x}_J, \mathbf{x}_L, \mathbf{x}_K, \mathbf{y}=\bm{1}_{S_1} + \bm{1}_{U_1} + \bm{1}_{T_1}) , \notag \\
    =& 
    \sum_{S_1 + U_1 \subseteq R \setminus T} \pi_{R_1}(\Sigma) \, 
    \mathcal{N}(\mathbf{x}_K \mid \bm{\mu}_K + \Sigma_{KI} G^T \bm{1}_{R_1}, \Sigma_{KK}).
    \label{eq:marginal}
\end{align}
In particular, when all binary variables are marginalized, the marginal distribution is precisely a $2^q$ mixture of Gaussian distributions with equal covariance, where mixing weights are given by $\pi_{R_1}(\Sigma)$ and the mean of the normal distributions is shifted by $\Sigma G^T \bm{1}_{R_1}$.
On the other hand, when all continuous variables are marginalized, the marginal distribution is no longer in the same form as the Grassmann distribution.

The conditional distribution with missing values for $\mathbf{x}_L$ and $\mathbf{y}_U$ is given by
\begin{align}
    p(\mathbf{x}_J, \mathbf{y}_S |\mathbf{x}_K, \mathbf{y}_T)  = &  \frac{\int d\mathbf{x}_{L} \sum_{U_1 \subseteq R \setminus (S + T)} p(\mathbf{x}_J, \mathbf{x}_L, \mathbf{x}_K, \mathbf{y}=\bm{1}_{S_1} + \bm{1}_{U_1} + \bm{1}_{T_1})}{\int d\mathbf{x}_{J}d\mathbf{x}_{L} \sum_{S_1' + U_1' \subseteq R \setminus T} p(\mathbf{x}_J, \mathbf{x}_L, \mathbf{x}_K, \mathbf{y}=\bm{1}_{S_1'} + \bm{1}_{U_1'} + \bm{1}_{T_1})} 
    = \frac{p(\mathbf{x}_J, \mathbf{x}_K, \mathbf{y}_S, \mathbf{y}_T)}{p(\mathbf{x}_K, \mathbf{y}_T)} , \notag \\
    =& 
    \frac{\sum_{U_1 \subseteq R \setminus (S + T)} \pi_{R_1}\bigl(\Sigma_{(J+L)|K}\bigr)
    e^{\bm{1}_{R_1}^T G \Sigma_{IK} \Sigma_{KK}^{-1} (\mathbf{x}_K - \bm{\mu}_K)}}{\sum_{S_1' + U_1' \subseteq R \setminus T} \pi_{R_1'}\bigl(\Sigma_{(J+L)|K}\bigr)
    e^{\bm{1}_{R_1'}^T G \Sigma_{IK} \Sigma_{KK}^{-1} (\mathbf{x}_K - \bm{\mu}_K)}} \notag \\
    & \hspace{-2cm}
    \mathcal{N}\bigl(\mathbf{x}_J \mid \bm{\mu}_J + \Sigma_{JK} \Sigma_{KK}^{-1} (\mathbf{x}_K -\bm{\mu}_K) + (\Sigma_{J (J+L)} - \Sigma_{JK} \Sigma_{KK}^{-1} \Sigma_{K(J+L)}) G_{(J+L) R}^T \bm{1}_{R_1}, \Sigma_{J|K} \bigr),
    \label{eq:conditional}
\end{align}
where $\Sigma_{KK}^{-1}$ denotes the inverse matrix of the submatrix $\Sigma_{KK}$ and the mixing weight is defined as previously mentioned,
\begin{align}
    \pi_{R_1}\bigl(\Sigma_{(J+L)|K}\bigr) \equiv
    \frac{\det (\Lambda_{R_0 R_0} -I) e^{\frac{1}{2} \bm{1}_{R_1}^T G_{R (J+L)} \Sigma_{(J+L)|K} G_{(J+L)R}^T \bm{1}_{R_1}}}{\sum_{R_1' \subseteq R} \det (\Lambda_{R_0' R_0'} -I) e^{\frac{1}{2} \bm{1}_{R_1'}^T G_{R (J+L)} \Sigma_{(J+L)|K} G_{(J+L)R}^T \bm{1}_{R_1'}}},
\end{align}
and $\Sigma_{J|K} \equiv \Sigma_{JJ} - \Sigma_{JK}\Sigma_{KK}^{-1}\Sigma_{KJ}$ is the Schur complement.

When there are no missing values, the conditional distribution is expressed more concisely:
\begin{align}
    p(\mathbf{x}_J, \mathbf{y}_S |\mathbf{x}_K=\mathbf{x}_{I \setminus J}, \mathbf{y}_T = \mathbf{y}_{R \setminus S}) 
    =& 
    \frac{ \pi_{R_1}(\Sigma_{J|K}) e^{ \bm{1}_{S_1}^T G \Sigma_{IK}\Sigma_{KK}^{-1} (\mathbf{x}_K -\bm{\mu}_K)}}{  \sum_{S_1' \subseteq R \setminus T} \pi_{R_1'}(\Sigma_{J|K}) e^{ \bm{1}_{S_1'}^T G \Sigma_{IK}\Sigma_{KK}^{-1} (\mathbf{x}_K -\bm{\mu}_K)}} \notag \\ 
    & \mathcal{N}\bigl(\mathbf{x}_J \mid \bm{\mu}_J + \Sigma_{JK} \Sigma_{KK}^{-1} (\mathbf{x}_K -\bm{\mu}_K) + \Sigma_{J|K} G_{JR}^T \bm{1}_{R_1}, \Sigma_{J|K} \bigr).
    \label{eq:conditional_distribution_without_marginal_variables}
\end{align}
In particular, when observed variables consist exclusively of binary variables, the conditional distribution is expressed as a normal distribution,
\begin{align}
    p(\mathbf{x}_J|\mathbf{x}_K=\mathbf{x}_{I \setminus J}, \mathbf{y}_R) =& 
    \mathcal{N} \left( \mathbf{x}_J \mid \bm{\mu}_J + \Sigma_{JK}\Sigma_{KK}^{-1} (\mathbf{x}_K -\bm{\mu}_K) + \Sigma_{J|K} G^T \mathbf{y} , \Sigma_{J|K}\right),
    \label{eq:conditional_distribution_continuous}
\end{align}
there, the mean of the distribution is shifted depending on the value of the binary variables conditioned.
On the other hand, when observed variables consist exclusively of continuous variables, the conditional distribution is expressed as a Grassmann distribution:
\begin{align}
    p(\mathbf{y}_S | \mathbf{x}_I, \mathbf{y}_T=\mathbf{y}_{R\setminus S}) = & \mathcal{G} \bigl(\mathbf{y}_S \mid I + (\Lambda-I)_{S|T_0} E^{- G_{SI} (\mathbf{x}_I - \bm{\mu}_I)} \bigr), \notag \\
    \equiv& 
    \frac{ \det \bigl[ (\Lambda-I)_{S_0|T_0} E^{- G_{S_0 I} (\mathbf{x}_I - \bm{\mu}_I)} \bigr] }{\det \bigl[I + (\Lambda - I)_{S|T_0} E^{-G_{SI}(\mathbf{x}_I - \bm{\mu}_I)}  \bigr]}, 
    \label{eq:conditional_distribution_binary} \\
    (\Lambda - I)_{S|T_0}  \equiv & \Lambda_{SS}-I - \Lambda_{ST_0}(\Lambda_{T_0T_0}-I)^{-1}\Lambda_{T_0 S},
\end{align}
where
\begin{align}
    E^{-G_{SI}(\mathbf{x}_I-\bm{\mu}_I)} \equiv & \mathrm{diag}(e^{-\mathbf{g}_s^T (\mathbf{x}_I-\bm{\mu}_I)}), \hspace{0.5cm} s \in S
\end{align}
is a diagonal matrix with non-negative diagonal elements.

\subsection{Interpretation of the interaction parameter}
In this subsection, we see that the parameter $G$ representing interaction between continuous and binary variables can be interpreted as a regression coefficient and a partial correlation coefficient.

First, let the partition of indices be $I=(J, L, K) = (j, \emptyset, K)$ and $R=(S, U, T) = (\emptyset, \emptyset, T)$ in the expression for the conditional distribution, Eq.~(\ref{eq:conditional_distribution_continuous}), where $\emptyset$ is the empty set.
Then, the linear combination of the covariates, $\eta_j$, in linear regression is given by
\begin{align}
    E[x_j|\mathbf{x}_K = \mathbf{x}_{I \setminus j}, \mathbf{y}_R] =& 
    \mu_j + \Sigma_{jK} \Sigma_{KK}^{-1}(\mathbf{x}_K-\bm{\mu}_K) + \Sigma_{jj|K} G_{jR}^T \, \mathbf{y}_R, \notag  \\
    =&
    \mu_j - \Lambda_{jj}^{-1} \Lambda_{jK}(\mathbf{x}_K-\bm{\mu}_K) + \Lambda_{jj}^{-1} G_{jR}^T \, \mathbf{y}_R, \notag \\
    \equiv & \eta_j \equiv g\bigl(E[x_j|\mathbf{x}_K, \mathbf{y}_R]\bigr),
\end{align}
where $g(\cdot)$ is a link function of the generalized linear model that relates the linear combination of the covariates and the mean of the objective variable $x_j$.
The above expression implies that the column vector of the matrix $G$, $[G]_{sj}, (s=1, 2, \dots, q)$, can be interpreted as a regression coefficient of the explanatory dummy variable $\mathbf{y}$, and thus supports the validity of the method of quantification in linear regression.

Next, let us consider the case of a binary objective variable.
We put the indices as $I=(J, L, K) = (\emptyset, \emptyset, K)$ and $R=(S, U, T) = (s, \emptyset, T)$ in the expression for the conditional distribution, Eq.~(\ref{eq:conditional_distribution_binary}).
Then, the conditional distribution becomes
\begin{align}
    p(y_s=1|\mathbf{x}_I, \mathbf{y}_T=\mathbf{y}_{R \setminus s}) = \frac{1}{1 + \left(\Lambda_{ss} - 1 - \Lambda_{sT_0} (\Lambda_{T_0 T_0} - I)^{-1} \Lambda_{T_0 s}\right) e^{ -G_{sI} (\mathbf{x}_I-\bm{\mu}_I)}}. 
\end{align}
When the explanatory variables consist exclusively of continuous variables, the above equation expresses the logistic regression, where the row vector of $G$, $[G]_{sj} \equiv [\mathbf{g}_{s}^T]_j, \; (j=1, 2, \dots, p)$, can be interpreted as a regression coefficient of the explanatory variables $\mathbf{x}_I$.
On the other hand, when the conditioning variables are a mixture of binary and continuous variables, the expression is no longer the same simple form as the logistic regression.

However, when the explanatory variables consist exclusively of binary variables, further consideration can be made.
In this case, the conditional distribution becomes
\begin{align}
    E[y_s|\mathbf{y}_T] =& p(y_s=1|\mathbf{y}_T = \mathbf{y}_{R \setminus s}), \notag \\
     =& \frac{1}{\Lambda_{ss} - \Lambda_{sT_0} (\Lambda_{T_0 T_0} - I)^{-1} \Lambda_{T_0 s}}.
\end{align}
Then, if we assume that the conditioning variables are conditionally independent of each other, i.e., $\Lambda_{TT}=\mathrm{diag}(\Lambda_{tt}), \; (t \in T)$, we obtain the following relation between the linear combination of the covariates $\eta_s$ and the mean of the objective variable $y_s$:
\begin{align}
    E[y_s|\mathbf{y}_T] =& \biggl[ \Lambda_{ss} - \sum_{t \in T} \frac{\Lambda_{st}\Lambda_{ts}}{\Lambda_{tt}-1} + \sum_{t \in T_1} \frac{\Lambda_{st}\Lambda_{ts}}{\Lambda_{tt}-1} \biggr]^{-1}, \notag \\
    \equiv & \biggl[ b_s + \sum_{t \in T} \frac{\Lambda_{st}\Lambda_{ts}}{\Lambda_{tt}-1} \, y_t \biggr]^{-1} = \frac{1}{\eta_s} \equiv g(\eta_s).
\end{align}
The above expression implies the binary regression with inverse link function.
In this case, the regression coefficient is proportional to $\Lambda_{st}\Lambda_{ts}$.
However, in general, i.e., if the explanatory variables are a mixture of continuous and binary variables or binary variables are correlated with each other, there is no justification for quantification.

To further discuss the interpretation of the parameter $G$, let us calculate the correlation between continuous and bianary variables.
First, naive mean and covariance of binary and continuous variables are calculated as follows:
\begin{align}
    E[\mathbf{y}] =& \sum_{R_1 \subseteq R} \pi_{R_1}(\Sigma) \bm{1}_{R_1} \equiv \bar{\mathbf{y}},  \\
    E[\mathbf{y} \mathbf{y}^T] =& \sum_{R_1 \subseteq R} \pi_{R_1}(\Sigma) \bm{1}_{R_1} \bm{1}_{R_1}^T \equiv T, \\
    \mathrm{Cov}[\mathbf{y}, \mathbf{y}^T] \equiv & E[\mathbf{y} \mathbf{y}^T] - E[\mathbf{y}] E[\mathbf{y}]^T, \notag \\
    =& T - \bar{\mathbf{y}} \bar{\mathbf{y}}^T,
\end{align}
\begin{align}
    E[\mathbf{x}] =& \bm{\mu} + \Sigma G^T \bar{\mathbf{y}} \equiv \bar{\mathbf{x}}, \\
    E[\mathbf{x} \mathbf{x}^T] =& \Sigma + \bm{\mu} \bm{\mu}^T + \bm{\mu} \bar{\mathbf{y}}^T G \Sigma + \Sigma G^T \bar{\mathbf{y}} \bm{\mu} + \Sigma G^T T G \Sigma, \notag \\
    =& \Sigma + \bar{\mathbf{x}} \bar{\mathbf{x}}^T + \Sigma G^T (T - \bar{\mathbf{y}} \bar{\mathbf{y}}^T) G \Sigma, \\
    \mathrm{Cov}[\mathbf{x}, \mathbf{x}^T] \equiv & E[\mathbf{x} \mathbf{x}^T] - E[\mathbf{x}] E[\mathbf{x}]^T, \notag \\
    =& \Sigma + \Sigma G^T \mathrm{Cov}[\mathbf{y}, \mathbf{y}^T] G \Sigma,
\end{align}
\begin{align}
    E[\mathbf{x} \mathbf{y}^T] =& \bm{\mu} \bar{\mathbf{y}}^T + \Sigma G^T T, \\
    \mathrm{Cov}[\mathbf{x}, \mathbf{y}^T] \equiv & E[\mathbf{x} \mathbf{y}^T] - E[\mathbf{x}] E[\mathbf{y}]^T, \notag \\
    =& \Sigma G^T \mathrm{Cov}[\mathbf{y}, \mathbf{y}^T].
\end{align}
When we define the correlation among variables as a Pearson correlation coefficient $\rho$, we obtain
\begin{align}
    \rho(x_j, x_k) \equiv & \frac{\mathrm{Cov}[x_j, x_k]}{\sqrt{\mathrm{Var}[x_j] \mathrm{Var}[x_k]}}, \notag \\
    =& 
    \frac{[\Sigma + \Sigma G^T (T - \bar{\mathbf{y}} \bar{\mathbf{y}}^T) G \Sigma]_{jk}}{\sqrt{[\Sigma + \Sigma G^T (T - \bar{\mathbf{y}} \bar{\mathbf{y}}^T) G \Sigma]_{jj} [\Sigma + \Sigma G^T (T - \bar{\mathbf{y}} \bar{\mathbf{y}}^T) G \Sigma]_{kk}}}, \\
    \rho(y_s, y_t) \equiv & \frac{\mathrm{Cov}[y_s, y_t]}{\sqrt{\mathrm{Var}[y_s] \mathrm{Var}[y_t]}}, \notag  \\
    =& 
    \frac{ [T - \bar{\mathbf{y}} \bar{\mathbf{y}}^T ]_{st}}{\sqrt{\bar{y}_s (1- \bar{y}_s) \bar{y}_t (1- \bar{y}_t)}},  \\
    \rho(x_j, y_s) \equiv & \frac{\mathrm{Cov}[x_j, y_s]}{\sqrt{\mathrm{Var}[x_j] \mathrm{Var}[y_s]}}, \notag  \\
    =& 
    \frac{[\Sigma G^T (T - \bar{\mathbf{y}} \bar{\mathbf{y}}^T)]_{js}}{\sqrt{[\Sigma + \Sigma G^T (T - \bar{\mathbf{y}} \bar{\mathbf{y}}^T) G \Sigma]_{jj} (T_{ss} - \bar{y}_s \bar{y}_s)}}.
\end{align}
Therefore, the parameter $G$ can not be interpreted as a naive correlation between continuous and binary variables.

Next, let us consider the partial correlation, the correlation by a conditional distribution.
We first calculate the partial correlation between continuous variables $x_j$ and $x_k$ given conditioning variables $\mathbf{x}_K$ and $\mathbf{y}_T$.
Let the partition of indices be $I=(J,L,K) = (j+k, \emptyset, K)$ and $R=(S,U,T) = (\emptyset, \emptyset, T)$ in the expression for the conditional distribution, Eq.~(\ref{eq:conditional_distribution_continuous}).
Since the conditional distribution is just a normal distribution in this case, the partial correlation is expressed as that of the normal distribution:
\begin{align}
    \rho(x_j, x_k |\mathbf{x}_K=\mathbf{x}_{I \setminus (j+k)}, \mathbf{y}_R)
    \equiv&  \frac{\mathrm{Cov}[x_j, x_k | \mathbf{x}_K, \mathbf{y}_R]}{\sqrt{\mathrm{Cov}[x_j, x_j | \mathbf{x}_K, \mathbf{y}_R] \mathrm{Cov}[x_k, x_k | \mathbf{x}_K, \mathbf{y}_R]}}, \notag \\
    =& \frac{[\Sigma_{(j+k)|K}]_{jk}}{\sqrt{[\Sigma_{(j+k)|K}]_{jj} [\Sigma_{(j+k)|K}]_{kk}}}.
\end{align}

The partial correlation between binary variables $y_s$ and $y_t$ is expressed as that of the Grassmann distribution.
We put the indices as $I=(J,L,K) = (\emptyset, \emptyset, K)$ and $R=(S,U,T) = (s+t, \emptyset, T=R \setminus (s + t))$ in the conditional distribution, Eq.~(\ref{eq:conditional_distribution_binary}).
Then, we can calculate the partial correlation as follows:
\begin{align}
    E[y_s|\mathbf{x}_I, \mathbf{y}_T = \mathbf{y}_{R \, \setminus \,  (s + t)}] =& \frac{ \bigl[ (\Lambda - I)_{(s+t)|T_0} \bigr]_{ss} e^{-\bm{1}_{s}^T G_{RI} (\mathbf{x}_I - \bm{\mu}_I)}}{\det \bigl[I +  (\Lambda - I)_{(s+t)|T_0} E^{-G_{(s+t) I} (\mathbf{x}_I - \bm{\mu}_I)} \bigr]}, \\
    E[y_s y_t|\mathbf{x}_I, \mathbf{y}_T = \mathbf{y}_{R \, \setminus \,  (s + t)}] =& \frac{1}{\det \bigl[I +  (\Lambda - I)_{(s+t)|T_0} E^{-G_{(s+t) I} (\mathbf{x}_I - \bm{\mu}_I)} \bigr]}, \\
    \mathrm{Cov}[y_s, y_t|\mathbf{x}_I, \mathbf{y}_T = \mathbf{y}_{R \, \setminus \,  (s + t)}] =& \frac{- \bigl[ (\Lambda - I)_{(s+t)|T_0} \bigr]_{st}  \bigl[ (\Lambda - I)_{(s+t)|T_0} \bigr]_{ts} 
    e^{- (\bm{1}_{s} + \bm{1}_{t})^T G_{RI} (\mathbf{x}_I - \bm{\mu}_I)}}{\det \bigl[I +  (\Lambda - I)_{(s+t)|T_0} E^{-G_{(s+t) I} (\mathbf{x}_I - \bm{\mu}_I)} \bigr]^2},  \\
    \rho(y_s, y_t |\mathbf{x}_I, \mathbf{y}_T = \mathbf{y}_{R \, \setminus \,  (s + t)}) = & \frac{\mathrm{Cov}[y_s, y_t|\mathbf{x}_I, \mathbf{y}_T = \mathbf{y}_{R \, \setminus \,  (s + t)}]}{\sqrt{\mathrm{Var}[y_s|\mathbf{x}_I, \mathbf{y}_T = \mathbf{y}_{R \, \setminus \,  (s + t)}]\mathrm{Var}[y_t|\mathbf{x}_I, \mathbf{y}_T = \mathbf{y}_{R \, \setminus \,  (s + t)}]}}.
\end{align}

Let us consider the partial correlation between binary and continuous variables $y_s$ and $x_j$.
We put the indices as $I=(J,L,K) = (j, \emptyset, K)$ and $R=(S,U,T) = (s, \emptyset, T)$ in the conditional distribution, Eq.~(\ref{eq:conditional_distribution_without_marginal_variables}).
Then, we obtain
\begin{align}
    p(x_j, y_s |\mathbf{x}_K = \mathbf{x}_{I \setminus j }, \mathbf{y}_T= \mathbf{y}_{R \setminus s})
    =&
    \frac{1}{1 + \frac{\pi_{T_1}(\Sigma_{jj|K})}{\pi_{(s_1 + T_1)}(\Sigma_{jj|K})} e^{- \bm{1}_{s_1}^T G \Sigma_{IK} \Sigma_{KK}^{-1} (\mathbf{x}_K - \bm{\mu}_K)}} \notag \\
    & \mathcal{N}\bigl( x_j  \mid \mu_j + \Sigma_{jK} \Sigma_{KK}^{-1} (\mathbf{x}_K -\bm{\mu}_K) + \Sigma_{jj|K} G_{jR}^T \bm{1}_{R_1}, \Sigma_{jj|K} \bigr), \\
    \pi_{T_1}(\Sigma_{jj|K}) \equiv & \frac{\det(\Lambda_{(s_0+T_0) (s_0+T_0)}-I) e^{\frac{1}{2} \bm{1}_{T_1}^T G_{Rj} \Sigma_{jj|K} G_{jR} \bm{1}_{T_1} }}{\sum_{R_1' \subseteq R} \det(\Lambda_{R_0' R_0'}-I) e^{\frac{1}{2} \bm{1}_{R_1'}^T G_{Rj} \Sigma_{jj|K} G_{jR} \bm{1}_{R_1'} }}.
\end{align}
Then, the mean and variance of the binary variable $y_s$ are calculated as those of the Bernoulli distribution:
\begin{align}
    E[y_s | \mathbf{x}_K, \mathbf{y}_T] =&
    \biggl[ 1 + \frac{\pi_{T_1}(\Sigma_{jj|K})}{\pi_{(s_1 + T_1)}(\Sigma_{jj|K})} e^{- \bm{1}_{s_1}^T G \Sigma_{IK} \Sigma_{KK}^{-1} (\mathbf{x}_K - \bm{\mu}_K)} \biggr]^{-1} , \\
    \mathrm{Var}[y_s | \mathbf{x}_K, \mathbf{y}_T] =& E[y_s | \mathbf{x}_K, \mathbf{y}_T] (1 - E[y_s | \mathbf{x}_K, \mathbf{y}_T]).
\end{align}
The mean and variance of the continuous variable $x_j$ are calculated as those of the Gaussian mixture model:
\begin{align}
    E[x_j | \mathbf{x}_K, \mathbf{y}_T] = & \mu_j + \Sigma_{jK} \Sigma_{KK}^{-1} (\mathbf{x}_K - \bm{\mu}_K) + \Sigma_{jj|K} G_{jR}^T \bm{1}_{T_1} + \Sigma_{jj|K} G_{js}^T E[y_s | \mathbf{x}_K, \mathbf{y}_T] , \\
    E[x_j^2 | \mathbf{x}_K, \mathbf{y}_T] =& \Sigma_{jj|K} + \bigl(1 - E[y_s | \mathbf{x}_K, \mathbf{y}_T]\bigr) \bigl(\mu_j + \Sigma_{jK}\Sigma_{KK}^{-1} (\mathbf{x}_K - \bm{\mu}_K) + \Sigma_{jj|K} G_{jR}^T \bm{1}_{T_1} \bigr)^2 \notag \\
    & +
    E[y_s | \mathbf{x}_K, \mathbf{y}_T] \bigl(\mu_j + \Sigma_{jK}\Sigma_{KK}^{-1} (\mathbf{x}_K - \bm{\mu}_K) + \Sigma_{jj|K} G_{jR}^T \bm{1}_{T_1} + \Sigma_{jj|K} G_{js}^T \bigr)^2 , \\
    \mathrm{Var}[x_j | \mathbf{x}_K, \mathbf{y}_T] \equiv & E[x_j^2 | \mathbf{x}_K, \mathbf{y}_T] - E[x_j | \mathbf{x}_K, \mathbf{y}_T]^2, \notag \\
    =& \Sigma_{jj|K} + \mathrm{Var}[y_s | \mathbf{x}_K, \mathbf{y}_T] (\Sigma_{jj|K} G_{js}^T)^2.
\end{align}
Using these expressions, the partial correlation between binary and continuous variables is calculated as follows:
\begin{align}
    E[x_j y_s | \mathbf{x}_K, \mathbf{y}_T] = & \bigl(\mu_j + \Sigma_{jK} \Sigma_{KK}^{-1} (\mathbf{x}_K - \bm{\mu}_K) + \Sigma_{jj|K} G_{jR}^T \bm{1}_{T_1} + \Sigma_{jj|K} G_{js}^T \bigr) E[y_s | \mathbf{x}_K, \mathbf{y}_T] , \\
    \mathrm{Cov}[x_j, y_s| \mathbf{x}_K, \mathbf{y}_T] \equiv & E[x_j y_s | \mathbf{x}_K, \mathbf{y}_T] - E[x_j | \mathbf{x}_K, \mathbf{y}_T] E[y_s | \mathbf{x}_K, \mathbf{y}_T], \notag \\
    =& \mathrm{Var}[y_s | \mathbf{x}_K, \mathbf{y}_T] \Sigma_{jj|K} G_{js}^T, \\
    \rho(x_j, y_s |\mathbf{x}_K, \mathbf{y}_T) \equiv & \frac{\mathrm{Var}[x_j, y_s| \mathbf{x}_K, \mathbf{y}_T]}{\sqrt{\mathrm{Var}[x_j | \mathbf{x}_K, \mathbf{y}_T] \mathrm{Var}[y_s | \mathbf{x}_K, \mathbf{y}_T]}} , \notag \\
    =& 
    \sqrt{\frac{\mathrm{Var}[y_s | \mathbf{x}_K, \mathbf{y}_T] }{ \Sigma_{jj|K} + \mathrm{Var}[y_s | \mathbf{x}_K, \mathbf{y}_T] (\Sigma_{jj|K} G_{js}^T)^2}} \Sigma_{jj|K} G_{js}^T.
\end{align}
The above expression implies that the parameter $G$ can be interpreted as the partial correlation between binary and continuous variables.

Lastly, we check that the proposed distribution successfully models the linear correlation between binary and continuous variables by analyzing a real dataset.
The data used in this analysis is the birth data from catdata package of R (programming language)~\cite{birth}.
The birth data contain information about the birth and pregnancy of 775 children that were born alive.
Features that can be considered as continuous variables include weight (in kilogram), height (in centimeter) and age of mother, weight (in gram) and height (in centimeter) of child, and length of pregnancy (in week), etc.
Features that can be considered as discrete variables include sex of child, the number of times the mother had been pregnant previously, the number of days the child spent in the intensive care unit, whether the pregnancy was a multiple birth, whether the child was born by Cesarean section, whether the birth was artificially induced, whether the membranes burst occurred before the beginning of the birth pangs ($\text{yes}=1, \text{no}=0$).
As a preprocessing for the analysis, the discrete variables were binarized by thresholding.
For example, if the number of times the mother had been pregnant previously is greater than zero, we encoded $\text{Previous}=1$; otherwise, $\text{Previous}=0$.
We also encoded $\text{Intensive}=1$ if the number of days the child spent in the intensive care unit is nonzero, and $\text{Intensive}=0$ otherwise.
We analyzed the data for non-twin children with a weight greater than $1500$, i.e., with a weight higher than infant with very low birth weight.

\begin{figure}[htbp]
    \centering
    \includegraphics[scale=1, pagebox=cropbox, clip]{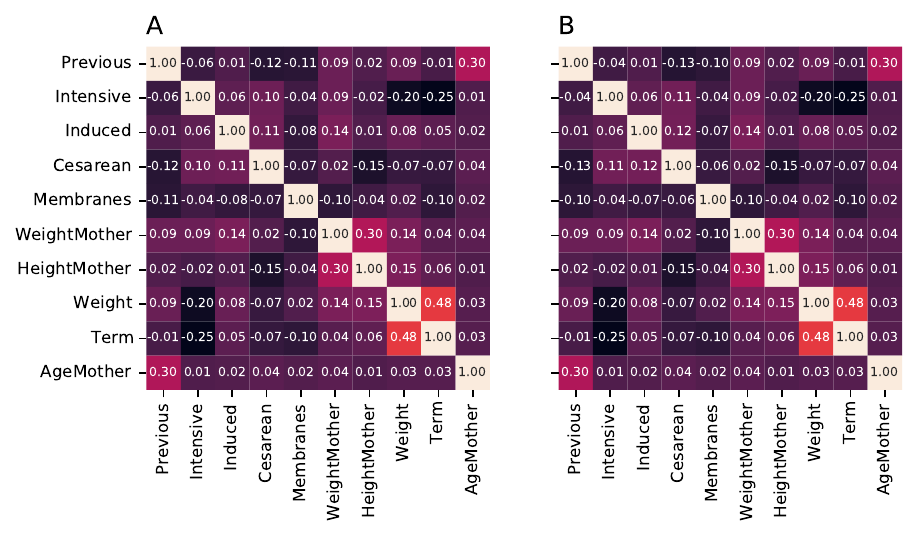}
    \caption{Pearson correlation matrix of the birth data, empirically computed from the data (A) and reproduced by the model (B).
    }
    \label{fig:correlation_matrix_birth}
\end{figure}

We sampled five binary features, birth experience (Previous), intensive care unit (Intensive), Cesarean, artificial induction (Induced), and membranes burst (Membranes), and five continuous features, weight and height and age of mother, weight of child, and length of pregnancy (Term), for analysis.
We used a diagonally dominant parameterization to ensure that $\Lambda - I$ is a $P_0$ matrix~\cite{Arai2021}.
Model parameters were estimated by maximum likelihood estimation.
The proposed distribution exactly reproduced the empirical mean of the data.
Fig.~\ref{fig:correlation_matrix_birth} represents the empirical correlation as well as correlation reproduced by the model.
We see that the proposed distribution successfully reproduces the correlation between binary and continuous variables.

\section{Factor analysis for a mixture of continuous and binary variables \label{sec:fa}}
In this section, we develop factor analysis for binary random variables as an application of the proposed distribution.
For convenience, we consider the case where observed variables are a mixture of binary and continuous variables.
Factor analysis is a model that expresses a correlation among observed variables through a continuous latent state in a lower dimensional space.
That is, observed binary and continuous variables can be compressed together into a continuous latent variable.
Of course, our model reduces to the usual factor analysis when binary variables are absent.

We denote the continuous and binary observed variables by $\mathbf{x}$ and $\mathbf{y}$ and denote continuous latent variables by $\mathbf{z}$.
Each variable is a column vector and its dimensions are $p_x$, $q$, and $p_z$, respectively.
We give the conditional distribution for the observed variables given the latent variable as a product of an uncorrelated normal distribution and an uncorrelated Bernoulli distribution with logit link function as follows:
\begin{align}
    p(\mathbf{x},\mathbf{y}|\mathbf{z}) = &p(\mathbf{x}|\mathbf{z}) p(\mathbf{y}|\mathbf{z}), \notag \\
    =&
    \mathcal{N}\bigl(\mathbf{x} \mid \bm{\mu}_x + W (\mathbf{z} - \bm{\mu}_z), \Psi\bigr)
    \prod_{j=1}^q \mathrm{Ber}\bigl(y_j \mid \mathrm{sigm}(b_{j} + \mathbf{g}_j^T (\mathbf{z}-\bm{\mu}_z)) \bigr), \notag \\
    \equiv &
    \frac{1}{(2\pi)^{p_x/2}\det\Psi^{1/2}} e^{-\frac{1}{2}(\mathbf{x}-\bm{\mu}_x-W (\mathbf{z} - \bm{\mu}_z))^T \Psi^{-1}(\mathbf{x}-\bm{\mu}_x-W (\mathbf{z} - \bm{\mu}_z))}
    \frac{e^{\mathbf{y}^T ( \mathbf{b} + G (\mathbf{z}-\bm{\mu}_z) )}}{\prod_{j=1}^q \bigl( 1 + e^{b_j + \mathbf{g}_j^T (\mathbf{z} -\bm{\mu}_z)} \bigr)},
    \label{eq:fa_conditional}
\end{align}
where $\mathrm{sigm}(\cdot)$ is a sigmoid function.
The above conditional distribution is parameterized by $(\bm{\mu}_x, \Psi, W)$ for continuous variables and $(\mathbf{b}, G)$ for binary variables.
The $p_x$-dimensional column vector $\bm{\mu}_x$ parameterizes the mean of observed continuous variables, and the $p_x \times p_x$ diagonal matrix $\Psi$ is a covariance matrix of observational noise.
The $p_x \times p_z$ matrix $W$ is a factor loading matrix for continuous variables~\cite{Murphy2012}.
The $q$-dimensional column vector $\mathbf{b}$ represents a bias term of the sigmoid function for binary variables, and the $q \times p_z$ matrix $G$ is a factor loading matrix for binary variables.
We give a prior distribution for $\mathbf{z}$ as a mixture of Gaussian distributions with equal covariance as follows:
\begin{align}
    p(\mathbf{z}) = & 
    \sum_{R_1 \subseteq R} \pi_{R_1}(\Sigma_z) \, \mathcal{N}(\mathbf{z} \mid \bm{\mu}_z + \Sigma_z G^T \bm{1}_{R_1}, \Sigma_z), \label{eq:fa_prior} \\
    \pi_{R_1}(\Sigma_z) \equiv & 
    \frac{e^{ \bm{1}_{R_{1}}^T \mathbf{b} + \frac{1}{2} \bm{1}_{R_{1}}^T G \Sigma_z G^T \bm{1}_{R_{1}}}}{\sum_{R_{1}' \subseteq R} e^{ \bm{1}_{R_{1}'}^T \mathbf{b} + \frac{1}{2} \bm{1}_{R_{1}'}^T G \Sigma_z G^T \bm{1}_{R_{1}'} }}.
\end{align}
Then, the observed distribution is induced as a continuous mixture of the conditional distribution:
\begin{align}
    p(\mathbf{x}, \mathbf{y}) =& \int_{-\infty}^{\infty} d\mathbf{z} \, p(\mathbf{x}, \mathbf{y}|\mathbf{z}) p(\mathbf{z}), \notag \\
    =&
    \pi_{R_1}(\Sigma_z) \, \mathcal{N}(\mathbf{x} \mid \bm{\mu}_x + W \Sigma_z G^T \mathbf{y}, \Sigma_x), \label{eq:fa_induced} \\
    \Sigma_x =& \Psi + W \Sigma_z W^T. 
\end{align}
When observed variables consist exclusively of binary variables, the observed distribution is exactly the same form as the Ising model~\cite{Ising1925}, where $\mathbf{b}$ is a bias term and $\frac{1}{2} G \Sigma_z G^T$ is a weight term of the Ising model.

A posterior distribution for $\mathbf{z}$ is simply given by a normal distribution:
\begin{align}
    p(\mathbf{z}|\mathbf{x}, \mathbf{y}) =& \mathcal{N}(\mathbf{z} \mid \mathbf{m}, \Sigma_{z|x}) , \label{eq:fa_posterior} \\
    \mathbf{m} =& \bm{\mu}_z + \Sigma_{z|x}[W^T \Psi^{-1}(\mathbf{x}-\bm{\mu}_x) + G^T \mathbf{y}] , \label{eq:factor_score} \\
    \Sigma_{z|x} =& [\Sigma_z^{-1} + W^T \Psi^{-1} W]^{-1}.
\end{align}
We call $\mathbf{m}$ in the above expression a factor score~\cite{Murphy2012}.
We can give a natural interpretation between prior and posterior distributions for $\mathbf{z}$; when we do not have observed variables, the prior distribution is given as a mixture of $2^q$ normal distributions since there is no information on the latent variable and it is uncertain.
By observing the binary variables, the posterior distribution reduces to one normal distribution out of $2^q$ mixture of normal distributions.
We give the expression for joint distribution for future reference:
\begin{align}
    p(\mathbf{x}, \mathbf{z}, \mathbf{y})
    =& 
    \pi_{R_1}(\Sigma_z) \, \mathcal{N}(\mathbf{x} \mid \bm{\mu}_x + W(\mathbf{z}-\bm{\mu}_z), \Psi)
    \, \mathcal{N}(\mathbf{z} \mid \bm{\mu}_z + \Sigma_z G^T \mathbf{y}, \Sigma_{z}). \label{eq:fa_joint}
\end{align}
More complete expressions for the proposed factor analysis with missing values are given in Appendix~\ref{sec:appendix_fa}.

The parameter $\Sigma_z$ can be renormalized to the redefinition of the parameters $G'=G \Sigma_z^{1/2}$ and $W'=W \Sigma_z^{1/2}$, and the scale transformation of the latent variable $\mathbf{z}'=\Sigma_z^{-1/2} \mathbf{z}$ and $\bm{\mu}_z' = \Sigma_z^{-1/2} \bm{\mu}_z $.
By contrast, the parameter $\bm{\mu}_z$ is irrelevant to the representability of the model and only affects the interpretation of the latent variable, since the likelihood function does not depend on $\bm{\mu}_z$.
Therefore, in the remainder of this paper, we set $\Sigma_z = I$ without loss of generality, and also set $\bm{\mu}_z = \bm{0}$.

As a factor analysis for binary variables, previous studies include binary Factor Analysis~\cite{Tipping1998}.
Binary Factor Analysis can be viewed as an example of a more general framework of exponential family PCA~\cite{Collins2001, Mohamed2008}, although mathematically it is more appropriate to call it ``factor analysis'' rather than ``principal component analysis''.
Exponential family PCA uses an exponential family distribution for the conditional distribution of observed variables given the latent variable $\mathbf{z}$.
For example, in binary Factor Analysis, the conditional distribution for observed variables is given by a Bernoulli distribution with the logit link function, $p(\mathbf{y}|\mathbf{z},\theta)= \prod_{j=1}^q \mathrm{Ber}(y_j \mid \mathrm{sigm}( w_0 + \mathbf{w}_j^T \mathbf{z}))$.
This choice of the conditional distribution is the same as in our model.
By contrast, the prior distribution for a latent variable is given by a normal distribution with zero mean and unit covariance, $p(\mathbf{z})=\mathcal{N}(\mathbf{z} \mid 0,I)$, unlike our model.
This introduction of the Gaussian prior distribution, however, has the disadvantage that the marginalization for the latent variable cannot be performed analytically, i.e., the induced distribution cannot be expressed analytically.
This drawback causes difficulty in estimating the parameters of the model.
Hence, one has to resort to an approximation technique for parameter estimation such as variational expectation-maximization algorithm~\cite{Tipping1998}, which approximates the functional form of the posterior distribution $p(\mathbf{z}|\mathbf{y})$, or Markov chain Monte Carlo (MCMC) simulation~\cite{Mohamed2008}, which is computationally demanding.
The difference between our model and previous studies ultimately lies in the introduction of a Gaussian mixture model as a prior distribution $p(\mathbf{z})$ for the latent variable.
This prior distribution allows us to perform the marginalization of the latent variable analytically.
This has the advantage that model parameters can be estimated by maximum likelihood estimation by using a common method such as gradient-based optimization.

\subsubsection{Improper solutions in maximum likelihood factor analysis}
It is quite important to mention the instability of model parameters in maximum likelihood estimation for usual factor analysis, although this instability has not been mentioned even in the standard textbooks~\cite{Bishop2006, Murphy2012}.
In the usual factor analysis, the covariance matrix of the induced distribution is given by the unique variance of observational noise $\Psi$ plus the contribution from the latent space $W W^T$:
\begin{align}
    p(\mathbf{x}) = \mathcal{N}\bigl(\mathbf{x} \mid \bm{\mu}_x, \Sigma_x = \Psi + W W^T \bigr).
\end{align}
At first glance, maximum likelihood estimation of the above expression with respect to the parameters $(\bm{\mu}_x, \Psi, W)$ seems to be no problem.
However, maximum likelihood estimates of the above expression are often unstable, e.g., the estimated value changes drastically as the number of latent dimensions varies or the dataset is changed slightly.
Specifically, some values of the diagonal elements of the observational noise covariance matrix $\Psi$ can become as close to zero as possible.
In some research fields, such a problem has been recognized as improper solutions~\cite{Gerbing1985, Gerbing1987}, or Heywood cases, of maximum likelihood estimation in factor analysis.
Although the causes of such instability have been investigated and various prescriptions for avoiding the instability have been proposed~\cite{Cooperman2022}, they are somewhat technical and do not seem to offer a fundamental solution.
We believe that this instability is the reason why probabilistic/non-probabilistic PCA has been preferred over factor analysis in practice~\cite{Price2006, Yamaguchi2008}, even though probabilistic PCA makes the unnatural assumption that the variance of observational noise for all observed variables is the same, i.e., homoscedastic.
That is, the results of probabilistic PCA depends on a scale transformation of observed variables.
We believe that this dependence on the scale transformation is not a desirable property for a data analysis method, even though in practice each observed variable is often standardized to roughly meet the homoscedasticity assumption.

We therefore propose a way to avoid the instability in maximum likelihood factor analysis.
We understand that the instability stems from too high degrees of freedom of the factor loading matrix $W$.
In fact, when each element of the parameter $W$ can take any values from each other, the parameter $W$ can reconstruct not only the correlation among observed variables but also the variance of observational noise.
In this case, particular diagonal elements of the observational noise covariance matrix $\Psi$ are close to zero, causing instability.
Therefore, to avoid the instability, we impose the constraint that the norm of the row vector of $\Psi^{-1/2} W$ is the same for all features, as follows:
\begin{align}
    \Sigma_x =& \Psi + W W^T, \notag \\
    =& \Psi + (c\Psi)^{1/2} \tilde{W} \tilde{W}^T (c\Psi)^{1/2}, \notag \\
    =& (1 + c) \Psi \; \biggl(1-\frac{c}{1 + c} \biggr) + (1+c) \Psi^{1/2} \tilde{W} \tilde{W}^T \Psi^{1/2} \; \biggl( \frac{c}{1 + c} \biggr), \notag \\
    =& \mathrm{diag}(W W^T) \frac{1}{c} + W W^T,
\end{align}
where the diagonal matrix $(1 + c) \Psi = \mathrm{diag}(\Sigma_x)$ expresses the diagonal elements of the induced distribution, and $\tilde{W}$ is a normalized factor loading matrix where each row vector is normalized to one.
The coefficient $c \ge 0$, which controls the strength of the influence of the latent space, is a squared norm of each row vector of a dimensionless factor loading matrix defined by $\Psi^{-1/2} W$.
The fraction $c/(1+c)$ represents the proportion of the variance of the observed variables that can be explained by the latent variable.
This norm constraint on the factor loading matrix $W$ allows for a clear distinction in the role of the parameters: $W$ is used exclusively to reconstruct the covariance among observed variables, while the covariance matrix $\Psi$ is used exclusively to account for the variance of observational noise.

As in the usual PCA, we can define the contribution ratio of the latent space in factor analysis by a reconstruction error of the variance of observed variables.
From the conditional distribution for the observed variables, Eq.~(\ref{eq:fa_conditional}), we see the linear relation between the linear combination of inputs $\bm{\eta}_x \equiv g(E[\mathbf{x}|\mathbf{z}]) = E[\mathbf{x}|\mathbf{z}] = \bm{\mu}_x + W \mathbf{z}$ and the latent variable $\mathbf{z}$:
\begin{align}
    \frac{1}{\sqrt{c \Psi}} (\bm{\eta}_x - \bm{\mu}_x) =& \tilde{W} \mathbf{z},
\end{align}
where $g(\cdot)$ is the link function and we have standardized the linear combination of inputs $\bm{\eta}_x$ by the net variance of $\mathbf{x}$, i.e., the total variance minus the variance comming from observational noise $ (1 + c) \Psi - \Psi = c\Psi$.
From this linear relation, we see that the variance of the linear combination of inputs can be expressed using the normalized factor loading matrix $\tilde{W}$:
\begin{align}
    \mathrm{Var}\bigl[ (c \Psi)^{-\frac{1}{2}} (\bm{\eta}_x - \bm{\mu}_x)\bigr]
    = & \tilde{W} \mathrm{Var}[\mathbf{z}] \tilde{W}^T, \notag \\
    =& \tilde{W} \tilde{W}^T
    = \sum_{s=1}^{p_z} \lambda_s \mathbf{u}_s \mathbf{u}_s^T,
\end{align}
where $\lambda_s$ and $\mathbf{u}_s$ are eigenvalues and eigenvectors of the matrix $\tilde{W} \tilde{W}^T$, respectively.
We see that the eigenvalue $\lambda_s$ represents the weight of each axis of the latent space in reconstructing the variation of the linear combination of inputs $\bm{\eta}_x$.
Therefore, the eigenvalues $\lambda_s$ can be used to define the contribution and cumulative contribution ratio of the latent space $P_s$ and $C_s$, respectively, as
\begin{align}
    P_s \equiv \frac{\lambda_s}{\sum_{t=1}^{p_z} \lambda_t}, \hspace{1cm}
    C_s \equiv \sum_{t=1}^s P_t, \label{eq:contribution_ratio}
\end{align}
where the axis of the latent space, the principal component axis, is sorted in descending order of the contribution ratio $P_s$.

We found that the proposed factor analysis for binary variables also suffers from the instability of model parameters similar to that of the continuous variables.
Hence, as in the case of continuous variables, we impose the constraint on the binary factor loading matrix $G$ that each row vector of $G$ has the same norm $c^{1/2}$ :
\begin{align}
    G = & c^{1/2} \, \tilde{G},
\end{align}
where each row vector of the normalized factor loading matrix $\tilde{G}$ is normalized to one.
In the case of factor analysis of binary variables, we see from Eq.~(\ref{eq:fa_conditional}) that the linear combination of inputs $\bm{\eta}_y = g(E[\mathbf{y}|\mathbf{z}]) = \mathrm{logit}(E[\mathbf{y}|\mathbf{z}]) =  \mathbf{b} + G \mathbf{z}$ has a linear relation to the latent variable $\mathbf{z}$ as
\begin{align}
    \frac{1}{\sqrt{c}} (\bm{\eta}_y - \mathbf{b})=&  \tilde{G} \mathbf{z}.
\end{align}
By an analogy from factor analysis for continuous variables, we propose to define the contribution ratio of the latent space by the eigenvalues of $\tilde{G} \tilde{G}^T$.
In other words, in the proposed binary factor analysis, the contribution ratio of the latent space is defined by the weight of each axis of the latent space in reconstructing the variation of the standardized linear combination of inputs.

In the case of factor analysis for a mixture of continuous and binary variables, the same norm constraint prescription as in the case of continuous and binary variables is applied to avoid the instability of model parameters in maximum likelihood estimation.
First, we express the model parameters using normalized factor loading matrices:
\begin{align}
    W = & (c \Psi)^{1/2} \, \tilde{W}, \notag \\
    G =& c^{1/2} \, \tilde{G},
\end{align}
where the coefficient $c$, representing the strength of the influence of the latent space, takes a common value for continuous and binary variables.
We define the combined normalized factor loading matrix $M$ as
\begin{align}
    M =
    \begin{bmatrix}
        \tilde{W} \\
        \tilde{G}
    \end{bmatrix}, \hspace{0.5cm} 
    \tilde{W} \equiv \begin{bmatrix} \;\;\; \tilde{\mathbf{w}}_1^T \;\;\;\;\; \\  \tilde{\mathbf{w}}_2^T \\ \vdots \\ \tilde{\mathbf{w}}_{p_x}^T \end{bmatrix}
    , \hspace{0.5cm} 
    \tilde{G} \equiv \begin{bmatrix} \;\;\; \tilde{\mathbf{g}}_1^T \;\;\;\;\; \\  \tilde{\mathbf{g}}_2^T \\ \vdots \\ \tilde{\mathbf{g}}_q^T \end{bmatrix}. \label{eq:factor_loading_vector}
\end{align}
Here, we shall call each row vector of normalized factor loading matrices, $\tilde{\mathbf{w}}_j^T$ and $\tilde{\mathbf{g}}_j^T$, a normalized factor loading vector.
The contribution ratio of the latent space is then defined by the eigenvalues of the matrix $M M^T$.

Finally, let us mention the identifiability of the model parameters.
As in the case of the usual factor analysis, the factor loading matrices $G$ and $W$ have rotational and sign reversal symmetry on the latent space.
In fact, the likelihood function is invariant under the rotational transformation of the combined factor loading matrix $M' = M R$, where $R$ is a rotation matrix.
Hence, we propose to fix the rotational degrees of freedom so that each column vector of the combined factor loading matrix $M'$ is orthogonal.
This orthogonality condition can be expressed as
\begin{align}
    (M')^T M' = \mathrm{diag}(\bm{\lambda}), \hspace{0.5cm} \bm{\lambda} = (\lambda_1, \lambda_2, \dots, \lambda_{p_z})^T, 
\end{align}
where $\lambda_s$ is an eigenvalue of the matrix $M^T M$.
The rotation matrix $R$ to satisfy the above orthogonality condition can be constructed by arranging the eigenvectors in columns, $R=[\mathbf{v}_1, \mathbf{v}_2, \dots, \mathbf{v}_{p_z}]$, where $\mathbf{v}_s$ is a column eigenvector of the matrix $M^T M$.

\subsection{Application of Factor Analysis to real datasets}
In this subsection, we numerically validate the proposed factor analysis and the norm constraint prescription using publicly available real datasets.

\subsubsection{HIV Drug Resistance Data}
In this section, we analyze the mutation of amino acid sequences of Human Immunodeficiency Virus (HIV) type-1.
The dataset was obtained from the HIV Drug Resistance Database published by Stanford University~\cite{hiv}.
Details of the database and related datasets can be found in Ref.~\cite{Rhee2003}.
When an antiretroviral drug is dosed on a patient, the virus becomes resistant to the drug over time by mutating its genes.
This mutation has been observed to be highly cooperative, with each residue of the amino acid sequence mutating not simply stochastically~\cite{Ohtaka2003}.
Although the molecular mechanism of drug resistance has not yet been elucidated, it is expected that the relationship between the correlation pattern of mutations and drug resistance will provide clues to the molecular mechanism of drug resistance.
We focused on viral resistance to protease inhibitors.
The data for analysis consists of mutational information on residues of amino acid sequences from position 1 to 99 in protease of viruses isolated from plasma of HIV-1 infected patients, represented by P1 to P99, and in vitro susceptibility to various protease inhibitors such as Nelfinavir.
As a preprocessing, the residues of amino acid sequences were encoded to $1$ if any mutation, such as insertion, deletion, or substitution from the consensus wild-type amino acid sequence, is present, and encoded to $0$ if there is no mutation from the consensus sequence.
In other words, information on the type of mutation was ignored.
To reduce the computational complexity, 10 out of the 99 residues were selected in descending order of mutation rate.
We used only the data without missing values in the mutations and the drug susceptibility.
After eliminating data with missing values, the sample size of the dataset was $N=2121$.

\begin{figure}[htbp]
    \centering
    \includegraphics[scale=1, pagebox=cropbox, clip]{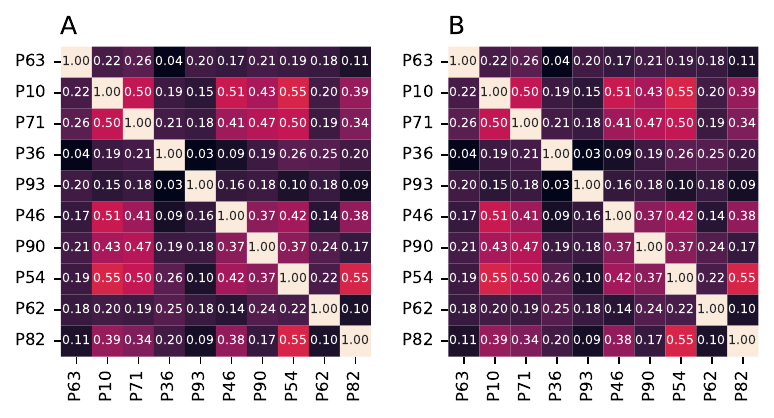}
    \caption{
    Pearson correlation matrix of the mutation of the amino acid sequences in HIV-1 protease, calculated from data (A) and reproduced by the proposed factor analysis (B).
    }
    \label{fig:correlation_matrix_hiv}
\end{figure}

We performed the proposed factor analysis on the mutation data of the amino acid sequences in HIV-1 protease.
Model parameters were estimated by maximum likelihood estimation.
The Bayesian information criterion (BIC) was used to determine the number of latent dimensions~\cite{BIC}.
The number of latent dimensions selected was $4$.
We also performed factor analysis in conjunction with the method of quantification for comparison, in which the number of latent dimensions was set to the same value as that of the proposed factor analysis and the norm constraint prescription was applied in a similar way.
The numerical analysis demonstrated that our factor analysis can exactly reproduce the empirical mean of the data.
Fig.~\ref{fig:correlation_matrix_hiv} shows the empirical correlation matrix and correlation reproduced by the model.
We see that the model successfully reproduces the empirical correlation.

Factor analysis allows us to visualize the relationship between data points and features, which is known as a biplot.
In the biplot, the factor scores of each data point $\mathbf{m}_i, \; (i=1,2,\dots, N)$, Eq.~(\ref{eq:factor_score}), are depicted as a scatter plot in Euclidean space, and the dimensionless factor loading vectors, which are defined by $c^{1/2} \, \tilde{\mathbf{w}}_j$ and $\mathbf{g}_j = c^{1/2} \, \tilde{\mathbf{g}}_j$, are depicted as arrows.
When the observed variables consist exclusively of binary variables, the factor score consists of $2^q$ possible combinations of the factor loading vectors $\mathbf{g}_j$.
The Euclidean distance between two data points represents the similarity between them.
The larger the inner product of the arrows of two features, the higher the similarity between them.
The inner product of data point $\mathbf{m}_i$ and the dimensionless factor loading vector $c^{1/2} \, \tilde{\mathbf{w}}_j$ and $\mathbf{g}_j$ also means that the corresponding feature is relatively larger or more likely to occur than mean value at that data point.
Our norm constraint prescription is also convenient in comparing features with each other in the biplot.
In the biplot, even if the orientation of normalized factor loading vectors of two features is similar, the similarity of these two features can be small when these lengths are different, i.e., the inner product is small.

\begin{figure}[htbp]
    \begin{center}
        \includegraphics[scale=1, pagebox=cropbox, clip]{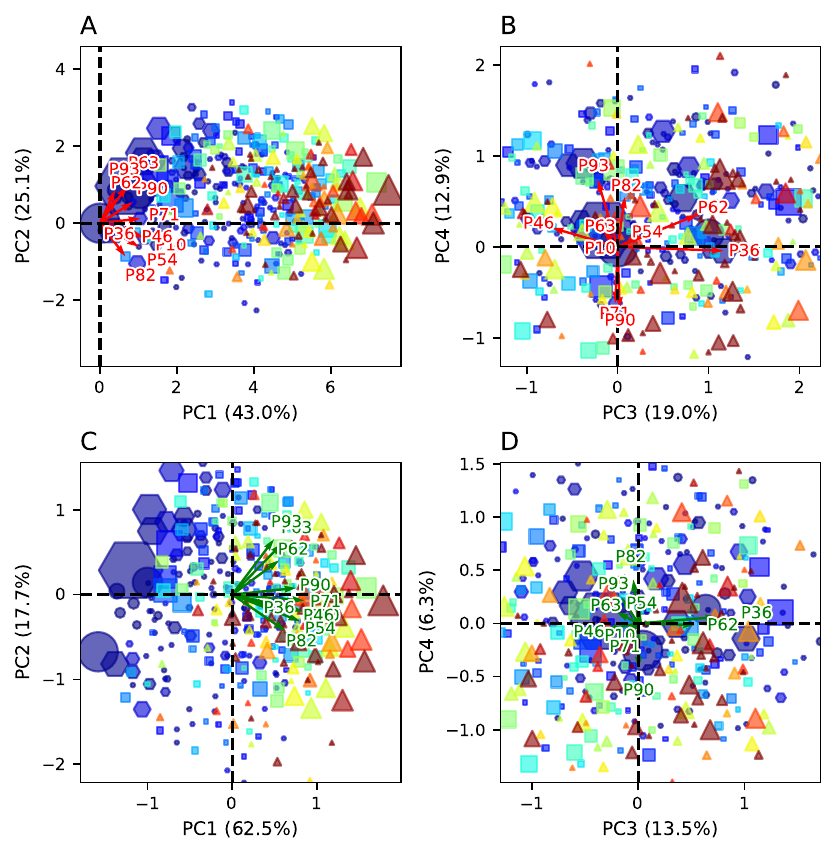}
    \end{center}
    \caption{
    Biplots of HIV-1 protease mutation data by the proposed factor analysis (A, B) and by the usual factor analysis with quantification (C, D).
    The point characters in the scatterplot have been changed depending on the resistance to the protease inhibitor (Nelfinavir): the circles denote the data with drug resistance smaller than the first quartile point, the hexagons are between the first and second quartile points, the squares are between the second and third quartile points, and the triangles are larger than the third quartile point.
    Those markers are colored from blue to red according to weak to strong drug resistance.
    The areas of the points in the scatterplot are proportional to the sample size of the corresponding data.
    }
    \label{fig:biplot_hiv}
\end{figure}

Biplots of factor analysis are shown in Fig.~\ref{fig:biplot_hiv}.
The axes of latent dimensions are displayed from the first principal component axis (PC1) to the fourth principal component axis (PC4), and the percentages in the axis labels represent the contribution ratio, Eq.~(\ref{eq:contribution_ratio}).
For comparison, the biplots of the usual factor analysis by the method of quantification with norm constraint are also shown.
From the biplot, we see that the first principal component axis (PC1) can be interpreted as the resistance to the protease inhibitor, and the second and subsequent principal component axes appear to be irrelevant to drug resistance.
We also see qualitative similarities between the proposed factor analysis and factor analysis with the method of quantification.
However, these two methods showed quantitative differences.
In particular, the contribution ratio of the principal component axis was quite different between these methods.
Proposed factor analysis appears to correctly reflect the contribution ratio of the latent dimension in the factor scores compared to factor analysis with quantification; in the proposed method, the data points of factor scores are spread more widely along the axis with a larger contribution ratio, while in the quantification method, the data points are uniformly spread along all axes.

\subsubsection{Birth data}
\begin{figure}[htbp]
    \centering
    \includegraphics[scale=1, pagebox=cropbox, clip]{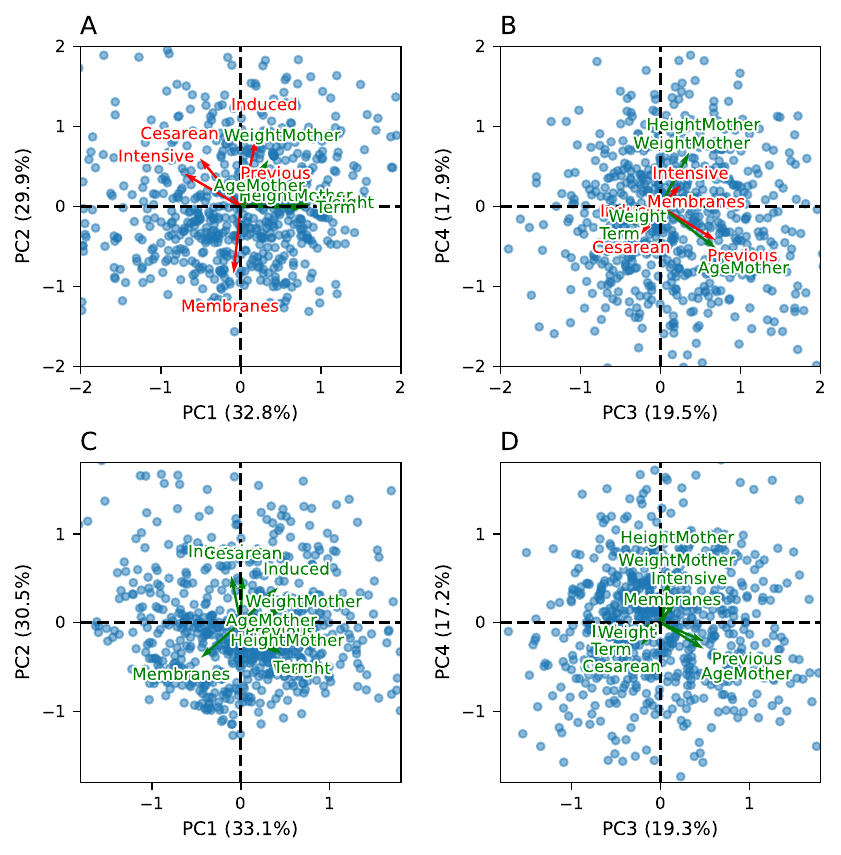}
    \caption{
    Biplots of birth data by the proposed factor analysis (A, B) and by the usual factor analysis with quantification (C, D).
    Factor loading vectors for variables treated as binary variables are represented by red arrows, while factor loading vectors for variables treated as continuous variables are represented by green arrows.
    }
    \label{fig:biplot_birth}
\end{figure}

To test our factor analysis with a mixture of continuous and binary variables, we analyzed the birth data used in the previous section.
Again, BIC was used to determine the number of latent dimensions.
The number of latent dimensions selected was $4$.
Again, we also performed factor analysis in conjunction with the method of quantification for comparison, where the norm constraint prescription was applied in a similar way.
The numerical analysis demonstrated that the model successfully reproduces the empirical mean and correlation of the data, as in the case of the HIV data.
Fig.~\ref{fig:biplot_birth} shows the biplot for the birth data.
The axes of latent dimensions are displayed from the first principal component axis (PC1) to the fourth principal component axis (PC4), and the percentages in the axis labels represent the contribution ratio, Eq.~(\ref{eq:contribution_ratio}). 
Unlike the HIV case, the proposed factor analysis showed even quantitatively similar results to that of the method of quantification.

\section{Conclusion \label{sec:conclusion}}
We proposed a multivariate probability distribution that models linear correlation between binary and continuous variables.
The proposed distribution essentially comprises $2^q$ normal distributions with equal variance and mean shifted by binary variables, where $q$ is the dimension of binary variables.
In the proposed distribution, the conditional distribution is expressed as the normal and Grassmann distribution while the marginal distribution is expressed as a mixture distribution.
As an application of the proposed distribution, we developed factor analysis for a mixture of continuous and binary observed variables.
We found that when observed variables consist exclusively of binary variables in the proposed factor analysis, the induced distribution on the observed variables can be expressed as the Ising model.
As an existing model of factor analysis for binary variables, there exists exponential family PCA~\cite{Tipping1998, Collins2001, Mohamed2008}.
Our model only differs from exponential family PCA in that our model introduces a Gaussian mixture model for a prior distribution over the latent variable, which allows us to express the induced distribution analytically.
Hence, our model has the advantage that model parameters can be estimated by maximum likelihood estimation by using a common method such as gradient-based optimization, whereas in the exponential family PCA, one has to resort to an approximation technique such as variational method, or a time-consuming Monte Carlo simulation such as MCMC methods.

We also discussed the instability of model parameters associated with maximum likelihood estimation in factor analysis, which is a problem known as improper solutions or Heywood cases in the literature~\cite{Gerbing1985, Gerbing1987}.
The proposed factor analysis also suffers from the improper solutions.
We empirically identify the cause of the improper solutions with the overlapping roles of the factor loading matrix $W$ and the covariance matrix of observational noise $\Psi$.
We then proposed a prescription to fix this instability, which imposes the constraint that the row vectors of the factor loading matrix have the same norm for all features.
We numerically confirmed by analyzing real datasets that this norm constraint prescription works well and avoids instability.

Since the proposed probability distribution successfully models the linear correlations between continuous and binary variables, the distribution can be used to develop many statistical learning methods such as clustering and anomaly detection.
Our factor analysis is also useful as a preprocessing for dimensional reduction and denoising of features and explanatory variables, which has conventionally been done with PCA.
For example, support vector machines, well-known methods for classification and regression, often use PCA as a preprocessing for explanatory variables before being applied.
However, when a dataset has binary features, the justification of applying PCA to such dataset is debatable, since PCA is originally devised to handle continuous features.
Our binary factor analysis is expected to be a useful alternative to conventional preprocessing methods when the dataset consists of a mixture of continuous and binary variables.
The proposed distribution is practically inapplicable to higher-dimensional binary data since one has to sum over all possible $2^q$ states to calculate the partition function.
Therefore, constructing a method that is applicable to higher-dimensional binary data is also a future work.

\section{Data Availability Statements}
The birth data underlying this article are publicly available from the catdata package of R language, which is published under the license of GPL-2~\cite{birth}.
The HIV data underlying this article are also publicly available from HIV Drug Resistance Database~\cite{hiv}.
Details of the database and related datasets are described in Ref.~\cite{Rhee2003}.

\appendix

\section{Theoretical background of the proposed model \label{sec:appendix}}
In this appendix, we introduce an interaction between continuous and binary variables based on the formulation with Grassmann numbers.
The definition and properties of Grassmann numbers can be consulted in Ref.~\cite{Arai2021}.
Readers who are interested in an application of the model rather than the theoretical background can safely skip this appendix.

\subsection{Introducing interaction between continuous and binary variables}
Let $\mathbf{x}$ denote a column vector of continuous variables and $\mathbf{y}$ denotes a column vector of binary variables, and let $p$ and $q$ denote their respective dimensions.
Assume that the binary variables are encoded as dummy variables taking the value $0$ or $1$.
That is, the vector $\mathbf{y}$ is a bit vector with each element taking the value $0$ or $1$.
We first consider the case where the binary variables are uncorrelated to the continuous variables.
We introduce a pair of $q$-dimensional vectors of Grassmann variables $(\bm{\theta}, \bar{\bm{\theta}})$, where $\bm{\theta}^{\dagger} \equiv \bar{\bm{\theta}}^T$.
Then, we define the Hamiltonian $H$ for calculating the expected value of the random variables as follows~\cite{Arai2021}:
\begin{align}
    \frac{1}{Z} e^{H} \equiv \frac{1}{Z} \, e^{-\frac{1}{2} (\mathbf{x} - \bm{\mu})^T \Sigma^{-1} (\mathbf{x} - \bm{\mu}) } \, e^{ \bm{\theta}^{\dagger} [I + (\Lambda-I)] \bm{\theta}},
\end{align}
where the partition function $Z$ is given by
\begin{align}
    Z = (2 \pi)^{p/2} \det \Sigma^{1/2} \det\Lambda.
\end{align}
The parameters $\bm{\mu}$ and $\Sigma$ are a column vector and square matrix with dimension $p$ representing the mean and covariance of the continuous variables, respectively.
$\Lambda - I$, where $I$ is an identity matrix, is a $P_0$ matrix with dimension $q \times q$.
An expected value can be calculated by the integral, $\int d\mathbf{x} \int d\bm{\theta} d\bar{\bm{\theta}}$, weighted by the Hamiltonian, e.g.,
\begin{align}
    E[x_j \, y_s] =& \int_{-\infty}^{\infty} d\mathbf{x} \int d\bm{\theta} d\bar{\bm{\theta}} \,
    \frac{1}{Z}  (x_j)(\bar{\theta}_s \theta_s) e^{-\frac{1}{2} (\mathbf{x} - \bm{\mu})^T \Sigma^{-1} (\mathbf{x} - \bm{\mu}) } \, e^{ \bm{\theta}^{\dagger} [I + (\Lambda-I)] \bm{\theta}} , \notag \\
    \equiv & \int_{-\infty}^{\infty} d\mathbf{x} \int \biggl[\prod_{r=1}^q d\theta_r d\bar{\theta}_r \biggr] 
    \frac{1}{Z} (x_j)(\bar{\theta}_s \theta_s) e^{-\frac{1}{2} (\mathbf{x} - \bm{\mu})^T \Sigma^{-1} (\mathbf{x} - \bm{\mu}) } \, e^{ \bm{\theta}^{\dagger} [I + (\Lambda-I)] \bm{\theta}}.
\end{align}
We denote the set of whole indices of continuous and binary variables as $I\equiv \{1,2,\dots,p\}$ and $R \equiv \{1,2,\dots, q\}$, respectively.
The set of whole indices for binary variables is divided into two parts with subscripts of $1$ or $0$, $R = (R_1, R_0)$, for variables that take the value $1$ or $0$, respectively.
Then, we write a subvector of the binary variables taking the value $1$ and $0$ as $\mathbf{y}_{R_1}=\bm{1}$ and $\mathbf{y}_{R_0}=\bm{0}$, respectively.
Noting that $p(y_s=0) = 1 - p(y_s=1)$, the joint probability $p(\mathbf{x}, \mathbf{y}_{R_1}=\bm{1}, \mathbf{y}_{R_0}=\bm{0})$ can be calculated as follows:
\begin{align}
    p(\mathbf{x},\mathbf{y}) =& p(\mathbf{x}, \mathbf{y}_{R_1}=\bm{1}, \mathbf{y}_{R_0}=\bm{0}), \notag \\
    =& 
    \frac{1}{Z} \int \biggl[ \prod_{r=1}^q d\theta_r d\bar{\theta}_i \biggr]
    \biggl[ \prod_{r_1 \in R_1} \bar{\theta}_{r_1} \theta_{r_1} \biggr]
    \biggl[ \prod_{r_0 \in R_0} (1 - \bar{\theta}_{r_0} \theta_{r_0})\biggr]
     e^{-\frac{1}{2} (\mathbf{x} - \bm{\mu})^T \Sigma^{-1} (\mathbf{x} - \bm{\mu}) } \, e^{ \bm{\theta}^{\dagger} [I + (\Lambda-I)] \bm{\theta}}, \notag \\
     \equiv & 
    \frac{1}{Z} \int d\bm{\theta} d\bar{\bm{\theta}} (\bar{\bm{\theta}}_{R_1} \bm{\theta}_{R_1}) (1-\bar{\bm{\theta}}_{R_0} \bm{\theta}_{R_0}) \, e^{-\frac{1}{2} (\mathbf{x} - \bm{\mu})^T \Sigma^{-1} (\mathbf{x} - \bm{\mu}) } \, e^{ \bm{\theta}^{\dagger} [I + (\Lambda-I)] \bm{\theta}}, \notag \\
    =& \frac{1}{(2\pi)^{p/2} \det \Sigma^{1/2}} \, e^{-\frac{1}{2}(\mathbf{x}-\bm{\mu})^T \Sigma^{-1} (\mathbf{x}-\bm{\mu})} \,
    \frac{1}{\det\Lambda} \int d\bm{\theta} d\bar{\bm{\theta}} (\bar{\bm{\theta}}_{R_1} \bm{\theta}_{R_1}) e^{-\bm{\theta}_{R_0}^{\dagger} \bm{\theta}_{R_0}} \, e^{\bm{\theta}^{\dagger} [I + (\Lambda-I)]\bm{\theta}}, \notag \\
    =&  \frac{1}{(2\pi)^{p/2} \det \Sigma^{1/2}} \, e^{-\frac{1}{2}(\mathbf{x}-\bm{\mu})^T \Sigma^{-1} (\mathbf{x}-\bm{\mu})} \,
    \frac{1}{\det\Lambda} \int d\bm{\theta}_{R_0} d\bar{\bm{\theta}}_{R_0} \, e^{\bm{\theta}_{R_0}^{\dagger} (\Lambda_{R_0 R_0}-I) \bm{\theta}_{R_0}}, \notag \\
    =& \frac{1}{(2\pi)^{p/2} \det \Sigma^{1/2}} \, e^{-\frac{1}{2}(\mathbf{x}-\bm{\mu})^T \Sigma^{-1} (\mathbf{x}-\bm{\mu})} \,
    \frac{1}{\det\Lambda} \, \det (\Lambda_{R_0 R_0} -I).
\end{align}

Extending the above formulation, let us introduce an interaction between binary and continuous variables.
We want to introduce the interaction in such a way that $\Lambda - I$ remains a $P_0$ matrix.
Since a $P_0$ matrix remains a $P_0$ matrix even when each column is multiplied by a positive constant~\cite{Tsatsomeros2002}, we can define the Hamiltonian which preserves the positivity of probability distribution as follows:
\begin{align}
    e^{H} \equiv e^{\bm{\theta}^{\dagger} [I + (\Lambda-I)  E^{- G (\mathbf{x}-\bm{\mu}) } ] \bm{\theta}} \,
    e^{-\frac{1}{2} (\mathbf{x} - \bm{\mu})^T \Sigma^{-1} (\mathbf{x} - \bm{\mu})} ,
\end{align}
where $G$ is a $q \times p$ matrix and $E^{- G (\mathbf{x}-\bm{\mu}) }$ is a diagonal matrix with positive diagonal elements:
\begin{align}
    G \equiv \begin{bmatrix} \;\;\; \mathbf{g}_1^T \;\;\;\;\; \\  \mathbf{g}_2^T \\ \vdots \\ \mathbf{g}_q^T \end{bmatrix}
    , \hspace{0.2cm} 
    E^{ - G (\mathbf{x} -\bm{\mu}) } \equiv &
    \begin{bmatrix}
        e^{- \mathbf{g}_1^T (\mathbf{x}-\bm{\mu})} & 0 & \cdots & 0 \\
        0 & e^{- \mathbf{g}_2^T (\mathbf{x}-\bm{\mu})} & \cdots & 0 \\
        \vdots & \vdots & \ddots  & \vdots \\
        0 &       0 &   \cdots &    e^{-\mathbf{g}_q^T (\mathbf{x}-\bm{\mu})}
    \end{bmatrix}, \notag \\
    = & \mathrm{diag}(e^{-\mathbf{g}_s^T (\mathbf{x}-\bm{\mu})}), \hspace{1cm} (s=1,2,\dots,q).
\end{align}
The partition function can be calculated by performing integration over the Grassmann variables $\bm{\theta}, \bar{\bm{\theta}}$ first:
\begin{align}
    Z \equiv & \int d\mathbf{x} d\bm{\theta} d\bar{\bm{\theta}} e^{H}, \notag \\
    =& \int d\mathbf{x} d\bm{\theta} d\bar{\bm{\theta}} \, e^{\bm{\theta}^{\dagger} [ I + (\Lambda-I) E^{- G  (\mathbf{x} - \bm{\mu}) } ] \bm{\theta}} \, e^{-\frac{1}{2} (\mathbf{x} - \bm{\mu} )^T \Sigma^{-1} (\mathbf{x} - \bm{\mu} )}, \notag \\
    =& \int d\mathbf{x} \, \det[I + (\Lambda - I) E^{- G (\mathbf{x} - \bm{\mu}) } ] \, e^{-\frac{1}{2} (\mathbf{x} - \bm{\mu} )^T \Sigma_x^{-1} (\mathbf{x} - \bm{\mu} )}, \notag \\
    =& \int d\mathbf{x} \, \sum_{R_0 \subseteq R} \det[ (\Lambda_{R_0 R_0} - I) \, E^{-G_{R_0 I}  (\mathbf{x} - \bm{\mu} ) }] \, e^{-\frac{1}{2}(\mathbf{x} - \bm{\mu} )^T \Sigma^{-1} (\mathbf{x} - \bm{\mu} )}, \notag \\
    =& \int d\mathbf{x} \sum_{R_0 \subseteq R} \det(\Lambda_{R_0 R_0} - I) e^{-\mathbf{g}_{R_0}^T (\mathbf{x}-\bm{\mu})} e^{-\frac{1}{2} (\mathbf{x}-\bm{\mu})^T \Sigma^{-1} (\mathbf{x}-\bm{\mu})}, \notag \\
    =& \int d\mathbf{x} \sum_{R_0 \subseteq R} \det(\Lambda_{R_0 R_0} - I) e^{-\frac{1}{2} (\mathbf{x}-\bm{\mu}+ \Sigma \mathbf{g}_{R_0})^T \Sigma^{-1} (\mathbf{x}- \bm{\mu} + \Sigma \mathbf{g}_{R_0})} e^{\frac{1}{2} \mathbf{g}_{R_0}^T \Sigma \mathbf{g}_{R_0}}, \notag \\
    =& (2\pi)^{p/2} \det \Sigma^{1/2} \sum_{R_0 \subseteq R} \det (\Lambda_{R_0 R_0}-I) e^{\frac{1}{2} \mathbf{g}_{R_0}^T \Sigma \mathbf{g}_{R_0}}, 
\end{align}
where we have defined a column vector $\mathbf{g}_{R_0} \equiv \sum_{s \in R_0} \mathbf{g}_s$, and the summation $\sum_{R_0 \subseteq R}$ runs over all possible principal minors.
That is, the partition function requires summing over all possible states for binary variables.
When we define a $q$-dimensional constant bit vector $\bm{1}_{R_0}$ whose elements take the value $0$ or $1$,
\begin{align}
    [\bm{1}_{R_0}]_s \equiv
    \begin{cases} 1, \hspace{0.5cm} \text{if} \; \; s \in R_0 \\ 
        0, \hspace{0.5cm} \text{if} \; \; s \in R_1
    \end{cases}, \; \; (s=1, 2, \dots, q),
    \label{eq:bit_vector_appendix}
\end{align}
the column vector $\mathbf{g}_{R_0}$ can also be expressed as
\begin{align}
    \mathbf{g}_{R_0} \equiv \sum_{s \in R_0} \mathbf{g}_s = G^T \bm{1}_{R_0}
    = \begin{bmatrix} \mathbf{g}_1, \mathbf{g}_2, \dots, \mathbf{g}_q \end{bmatrix} \bm{1}_{R_0}.
\end{align}

To express the joint, marginal and conditional distributions, we define the notation of index labels.
We denote the index label of a subset of whole indices as $J \subseteq I$.
The subvector consisting of the subset of indices $J$ is represented by $\mathbf{x}_J$.
We divide the set of whole indices of continuous and binary variables $I$ and $R$ into three subset parts; $I=(J, L, K)$ and $R=(S, U, T)$, where the index labels for a set of indices $L$ and $U$ are introduced to handle missing values.
Hence, the index labels $L$ and $U$ may be understood as the initial letters of ``Latent'' and ``Unobserved'', respectively.
The number of elements in the set of indices is represented by $p_J$, $p_L$, $p_K$ and $q_S$, $q_U$, $q_T$, these satisfy $p_J + p_L + p_K = p$ and $q_S + q_U + q_T = q$.
Then, the vectors $\mathbf{x}$ and $\mathbf{y}$ can be partitioned into subvectors as $\mathbf{x}=\mathbf{x}_I = (\mathbf{x}_J, \mathbf{x}_L, \mathbf{x}_K)$ and $\mathbf{y}=\mathbf{y}_R=(\mathbf{y}_S, \mathbf{y}_U, \mathbf{y}_T)$, respectively.
Again, a subset of indices for binary variables, e.g., $S$, is divided into two parts; we write the index label for the variable that takes the value $y_s=1$ and $y_s=0, \;\; (s \in S)$ as $S_1$ and $S_0$, and denote these variables as $\mathbf{y}_{S_1}$ and $\mathbf{y}_{S_0}$, respectively.
The union of the index label $J$ and $K$ is denoted as $J + K \equiv J \cup K$.

Using these notations, the joint distribution can be calculated as follows:
\begin{align}
    p(\mathbf{x},\mathbf{y}) =& p(\mathbf{x}, \mathbf{y}_{R_1}=\bm{1}, \mathbf{y}_{R_0}=\bm{0}) , \notag \\
    = &  \frac{1}{Z} \,  \int d\bm{\theta} d\bar{\bm{\theta}} \, (\bar{\bm{\theta}}_{R_1} \bm{\theta}_{R_1}) (1 - \bar{\bm{\theta}}_{R_0} \bm{\theta}_{R_0}) e^{\bm{\theta}^{\dagger}[I + (\Lambda-I) E^{- G  (\mathbf{x}-\bm{\mu})}]\bm{\theta}} \, e^{-\frac{1}{2}(\mathbf{x}-\bm{\mu})^T \Sigma^{-1}(\mathbf{x}-\bm{\mu})}, \notag \\
    =& \frac{1}{Z}  \int d\bm{\theta}_{R_0} d\bar{\bm{\theta}}_{R_0} e^{\bm{\theta}_{R_0}^{\dagger}(\Lambda_{R_0 R_0}-I)E^{- G_{R_0 I} (\mathbf{x}-\bm{\mu}) }\bm{\theta}_{R_0}} e^{-\frac{1}{2}(\mathbf{x}-\bm{\mu})^T \Sigma^{-1}(\mathbf{x}-\bm{\mu})}, \notag \\
    = & \frac{1}{Z} \det (\Lambda_{R_0 R_0}-I) \, e^{-\mathbf{g}_{R_0}^T (\mathbf{x}-\bm{\mu})} e^{-\frac{1}{2}(\mathbf{x}-\bm{\mu})^T \Sigma^{-1}(\mathbf{x}-\bm{\mu})}, \notag \\
    =& \frac{1}{Z} \det (\Lambda_{R_0 R_0} -I) \, e^{\frac{1}{2} \mathbf{g}_{R_0}^T \Sigma \mathbf{g}_{R_0}} e^{-\frac{1}{2}(\mathbf{x}-\bm{\mu} + \Sigma \mathbf{g}_{R_0})^T \Sigma^{-1}(\mathbf{x}-\bm{\mu} + \Sigma \mathbf{g}_{R_0})}.
\end{align}
By marginalizing the variables $\mathbf{x}_{J+L}$ and $\mathbf{y}_{S+U}$, the marginal distribution can be calculated as follows:
\begin{align}
    & p(\mathbf{x}_K, \mathbf{y}_T) = p(\mathbf{x}_K, \mathbf{y}_{T_1}=\bm{1}, \mathbf{y}_{T_0}=\bm{0}) = \int d\mathbf{x}_{J+L} \sum_{\mathbf{y}_{S+U} \in \{0, 1\}^{q_S + q_U}} p(\mathbf{x}, \mathbf{y}), \notag \\
    =& \frac{1}{Z} \int d\mathbf{x}_{J+L} d\bm{\theta} d\bar{\bm{\theta}} (\bar{\bm{\theta}}_{T_1} \bm{\theta}_{T_1}) (1 - \bar{\bm{\theta}}_{T_0} \bm{\theta}_{T_0}) e^{\bm{\theta}^{\dagger} [I + (\Lambda-I)E^{-G (\mathbf{x}-\bm{\mu})}] \bm{\theta}} e^{-\frac{1}{2}(\mathbf{x}-\bm{\mu})^T \Sigma^{-1} (\mathbf{x}-\bm{\mu})}, \notag \\
    =& \frac{1}{Z} \int d\mathbf{x}_{J+L} d\bm{\theta}_{S+U} d\bar{\bm{\theta}}_{S+U} d\bm{\theta}_{T_0} d\bar{\bm{\theta}}_{T_0} \notag \\
    & \hspace{-1.6cm}
    \exp\biggl\{ \begin{bmatrix} \bm{\theta}_{S+U}^{\dagger}, & \bm{\theta}_{T_0}^{\dagger} \end{bmatrix}
    \begin{bmatrix}
        I + (\Lambda_{(S+U)(S+U)}-I) E^{-G_{(S+U)I} (\mathbf{x}-\bm{\mu})} & \Lambda_{(S+U) T_0} E^{-G_{T_0 I} (\mathbf{x}-\bm{\mu})} \\
        \Lambda_{T_0 (S+U)} E^{-G_{(S+U) I} (\mathbf{x}-\bm{\mu})} & (\Lambda_{T_0 T_0}-I) E^{-G_{T_0 I}  (\mathbf{x}-\bm{\mu})}
    \end{bmatrix}
    \begin{bmatrix} \bm{\theta}_{S+U} \\ \bm{\theta}_{T_0} \end{bmatrix} \biggr\}
    e^{-\frac{1}{2}(\mathbf{x} -\bm{\mu})^T \Sigma^{-1} (\mathbf{x} -\bm{\mu})}, \notag \\
    =& \frac{1}{Z} \int d\mathbf{x}_{J+L} \det
        \begin{bmatrix}
            I + (\Lambda_{(S+U)(S+U)}-I) E^{-G_{(S+U)I} (\mathbf{x}-\bm{\mu})} & \Lambda_{(S+U) T_0} E^{-G_{T_0 I} (\mathbf{x}-\bm{\mu})} \\
            \Lambda_{T_0 (S+U)} E^{-G_{(S+U)I} (\mathbf{x}-\bm{\mu})} & (\Lambda_{T_0 T_0}-I) E^{-G_{T_0 I} (\mathbf{x}-\bm{\mu})}
        \end{bmatrix}
        e^{-\frac{1}{2}(\mathbf{x} -\bm{\mu})^T \Sigma^{-1} (\mathbf{x} -\bm{\mu})}, \notag \\
    =& \frac{1}{Z} \int d\mathbf{x}_{J+L} \det[(\Lambda_{T_0 T_0} - I) E^{-G_{T_0 I} (\mathbf{x}-\bm{\mu})}] \notag \\
    & \hspace{-1.4cm}
    \det \bigl[I + (\Lambda_{(S+U)(S+U)}-I) E^{-G_{(S+U) I} (\mathbf{x}-\bm{\mu})} - \Lambda_{(S+U) T_0} (\Lambda_{T_0 T_0}-I)^{-1} \Lambda_{T_0 (S+U)} E^{-G_{(S+U) I} (\mathbf{x}-\bm{\mu})} \bigr]
    e^{-\frac{1}{2} (\mathbf{x}-\bm{\mu})^T \Sigma^{-1}(\mathbf{x}-\bm{\mu})}, \notag \\
    =& \frac{1}{Z} \int d\mathbf{x}_{J+L} \sum_{S_0 \subseteq S, \; U_0 \subseteq U} \det(\Lambda_{R_0 R_0} - I) 
    e^{-\mathbf{g}_{R_0}^T (\mathbf{x}-\bm{\mu})} e^{-\frac{1}{2} (\mathbf{x}-\bm{\mu})^T \Sigma^{-1} (\mathbf{x}-\bm{\mu})}, \notag \\
    =& \frac{1}{Z} \int d\mathbf{x}_{J+L} \sum_{S_0 \subseteq S, \; U_0 \subseteq U} \det(\Lambda_{R_0 R_0} - I) 
    e^{\frac{1}{2} \mathbf{g}_{R_0}^T \Sigma \mathbf{g}_{R_0}} e^{-\frac{1}{2} (\mathbf{x} -\bm{\mu} + \Sigma \mathbf{g}_{R_0})^T \Sigma^{-1} (\mathbf{x} -\bm{\mu} + \Sigma \mathbf{g}_{R_0})}, \notag \\
    =& \frac{1}{Z} \int d\mathbf{x}_{J+L} \sum_{S_0 \subseteq S, \; U_0 \subseteq U} \det(\Lambda_{R_0 R_0} - I) 
    e^{\frac{1}{2} \mathbf{g}_{R_0}^T \Sigma \mathbf{g}_{R_0}}
    e^{-\frac{1}{2} (\mathbf{x}_K -\bm{\mu}_K + \Sigma_{KI} \mathbf{g}_{I R_0})^T \Sigma_{KK}^{-1} (\mathbf{x}_K -\bm{\mu}_K + \Sigma_{KI} \mathbf{g}_{I R_0})} \notag \\
    & \hspace{-1.8cm} e^{-\frac{1}{2} (\mathbf{x}_{J+L} - \bm{\mu}_{J+L} + \Sigma_{(J+L)I} \mathbf{g}_{R_0} - \Sigma_{(J+L) K} \Sigma_{KK}^{-1} (\mathbf{x}_K - \bm{\mu}_K + \Sigma_{KI} \mathbf{g}_{I R_0}))^T \Sigma_{(J+L)|K}^{-1} (\mathbf{x}_{J+L} - \bm{\mu}_{J+L} + \Sigma_{(J+L)I} \mathbf{g}_{R_0} - \Sigma_{(J+L) K} \Sigma_{KK}^{-1} (\mathbf{x}_K - \bm{\mu}_K + \Sigma_{KI} \mathbf{g}_{I R_0})) }, \notag  \\
    =& \frac{1}{Z} (2\pi)^{p_{J+L}/2} \det \Sigma_{(J+L)|K}^{1/2} \sum_{S_0 \subseteq S, \; U_0 \subseteq U} \det(\Lambda_{R_0 R_0} - I)  e^{\frac{1}{2} \mathbf{g}_{R_0}^T \Sigma \mathbf{g}_{R_0}}
    e^{-\frac{1}{2} (\mathbf{x}_K -\bm{\mu}_K + \Sigma_{KI} \mathbf{g}_{I R_0})^T \Sigma_{KK}^{-1} (\mathbf{x}_K -\bm{\mu}_K + \Sigma_{KI} \mathbf{g}_{I R_0})},
\end{align}
where we have defined the division of the vector $\mathbf{g}_{R_0}^T$ as $\mathbf{g}_{R_0}^T = \mathbf{g}_{R_0 I}^T = (\mathbf{g}_{R_0 J}^T, \mathbf{g}_{R_0 L}^T, \mathbf{g}_{R_0 K}^T)$.
The conditional distribution with marginalized variables can be derived by dividing marginal distributions:
\begin{align}
    & p(\mathbf{x}_J, \mathbf{y}_{S_1}=\bm{1}, \mathbf{y}_{S_0}=\bm{0} |\mathbf{x}_K, \mathbf{y}_{T_1}=\bm{1}, \mathbf{y}_{T_0}=\bm{0}) 
    = \frac{p(\mathbf{x}_{J+K}, \mathbf{y}_{S_1+T_1}=\bm{1}, \mathbf{y}_{S_0+T_0}=\bm{0})}{p(\mathbf{x}_K, \mathbf{y}_{T_1}=\bm{1}, \mathbf{y}_{T_0}=\bm{0})}, \notag \\
    =& 
    \frac{ \sum_{U_0 \subseteq U} \det(\Lambda_{R_0 R_0}-I)
    e^{\frac{1}{2} \mathbf{g}^T_{R_0 (J+L)} \Sigma_{(J+L)|K} \mathbf{g}_{(J+L) R_0}}
    e^{-\mathbf{g}_{(S_0+U_0) I}^T \Sigma_{IK} \Sigma_{KK}^{-1} (\mathbf{x}_K -\bm{\mu}_K)}}
    {\sum_{S_{0}' \subseteq S, \; U_{0}' \subseteq U} \det(\Lambda_{R_{0}' R_{0}'} - I)
    e^{\frac{1}{2} \mathbf{g}^T_{R_0' (J+L)} \Sigma_{(J+L)|K} \mathbf{g}_{(J+L) R_0'}}
    e^{-\mathbf{g}_{(S_{0}' + U_{0}') K}^T \Sigma_{IK} \Sigma_{KK}^{-1} (\mathbf{x}_K -\bm{\mu}_K)}  } \notag \\
    & \hspace{-1cm}
    \frac{1}{(2\pi)^{p_J/2} \det\Sigma_{J|K}^{1/2}} 
    e^{-\frac{1}{2}(\mathbf{x}_J - \bm{\mu}_J + \Sigma_{JI} \mathbf{g}_{IR_0} - \Sigma_{JK} \Sigma_{KK}^{-1} (\mathbf{x}_K - \bm{\mu}_K + \Sigma_{KI} \mathbf{g}_{IR_0}))^T \Sigma_{J|K}^{-1}(\mathbf{x}_J - \bm{\mu}_J + \Sigma_{JI} \mathbf{g}_{IR_0} - \Sigma_{JK} \Sigma_{KK}^{-1} (\mathbf{x}_K - \bm{\mu}_K + \Sigma_{KI} \mathbf{g}_{IR_0})) }.
\end{align}

The above expressions can also be expressed using a constant bit vector defined in Eq.~(\ref{eq:bit_vector_appendix}).
In fact, we have $\mathbf{g}_{R_0} = G^T \bm{1}_{R_0} = G^T \tilde{\bm{1}}_{R_1} \equiv G^T (\bm{1} - \bm{1}_{R_1})$.
Redefining the parameters as $\tilde{\bm{\mu}} \equiv \bm{\mu} - \Sigma G^T \bm{1}$ and $(\widetilde{\Lambda - I}) \equiv (\Lambda - I) E^{G \Sigma G^T \bm{1}}$, and using the dummy vector $\mathbf{y}$ explicitly, the joint, marginal and conditional distributions are also expressed as follows:
\begin{align}
    p(\mathbf{x}, \mathbf{y}=\bm{1}_{R_1}) 
    =& \pi_{R_1}(\Sigma) \, \mathcal{N}( \mathbf{x} \mid \tilde{\bm{\mu}} + G^T \mathbf{y}, \Sigma ), \\
    \pi_{R_1}(\Sigma)
    \equiv&
    \frac{\det \Bigl[(\widetilde{\Lambda -I})_{R_0 R_0} \Bigr]  e^{\frac{1}{2} \bm{1}_{R_1}^T G \Sigma G^T \bm{1}_{R_1}}}{\sum_{R_1' \subseteq R} \det \Bigl[(\widetilde{\Lambda-I})_{R_0' R_0'}\Bigr] e^{\frac{1}{2} \bm{1}_{R_1'}^T G \Sigma G^T \bm{1}_{R_1'}}},
\end{align}
\begin{align}
    p(\mathbf{x}_K, \mathbf{y}_T) 
    =& \sum_{S_1 + U_1 \subseteq R \setminus T} \pi_{R_1}(\Sigma) \,
    \mathcal{N}( \mathbf{x}_K \mid \tilde{\bm{\mu}}_K + \Sigma_{KI} G^T \bm{1}_{R_1}, \Sigma_{KK} ),
\end{align}
\begin{align}
    & p(\mathbf{x}_J, \mathbf{y}_S |\mathbf{x}_K, \mathbf{y}_T)  \notag \\
    =& \frac{\sum_{U_1 \subseteq R\setminus(S+T)} \det \Bigl[ (\widetilde{\Lambda-I})_{R_0 R_0} \Bigr]
    e^{\frac{1}{2} \bm{1}_{R_1}^T G_{R(J+L)} \Sigma_{(J+L)|K} G_{(J+L)R}^T \bm{1}_{R_1}}
    e^{(\bm{1}_{S_1} + \bm{1}_{U_1})^T G \Sigma_{IK} \Sigma_{KK}^{-1} (\mathbf{x}_K - \tilde{\bm{\mu}}_K)}}
    {\sum_{S_1' + U_1' \subseteq R \setminus T} \det\Bigl[ (\widetilde{\Lambda-I})_{R_0' R_0'} \Bigr]
    e^{\frac{1}{2} \bm{1}_{R_1'}^T G_{R(J+L)} \Sigma_{(J+L)|K} G_{(J+L)R}^T \bm{1}_{R_1'}}
    e^{(\bm{1}_{S_1'} + \bm{1}_{U_1'})^T G \Sigma_{IK} \Sigma_{KK}^{-1} (\mathbf{x}_K - \tilde{\bm{\mu}}_K)}} \notag \\
    &
    \mathcal{N}\bigl(\mathbf{x}_J \mid \tilde{\bm{\mu}}_J + \Sigma_{JI} G^T \bm{1}_{R_1} + \Sigma_{JK} \Sigma_{KK}^{-1}(\mathbf{x}_K - \tilde{\bm{\mu}}_K - \Sigma_{KI} G^T \bm{1}_{R_1}), \Sigma_{J|K} \bigr).
\end{align}
When we redefine the parameters as $\widetilde{\Lambda - I} \rightarrow \Lambda - I$ and $\tilde{\bm{\mu}} \rightarrow \bm{\mu}$, we obtain the expressions in the main text of the paper, Eqs.~(\ref{eq:joint}, \ref{eq:marginal}, \ref{eq:conditional}).

\subsection{Derivation of factor analysis for a mixture of continuous and binary variables \label{sec:appendix_fa}}
The proposed binary factor analysis can be realized as a special case of the proposed distribution.
In factor analysis, observed variables correlate through a continuous latent variable.
We denote the observed binary variables and continuous variables and latent variables as $\mathbf{y}$, $\mathbf{x}$, and $\mathbf{z}$, respectively, and their respective dimensions are $q$, $p_x$, and $p_z$.
In factor analysis, we assume that each binary variable $y_s$ is conditionally uncorrelated.
That is, $\Lambda$ is a diagonal matrix, $\Lambda - I = E^{- \mathbf{b}} \equiv \mathrm{diag}(e^{-b_s}), \hspace{0.2cm}\widetilde{\Lambda - I} = E^{- \tilde{\mathbf{b}}} \equiv \mathrm{diag}(e^{-\tilde{b}_s}), \hspace{0.2cm} (s=1,2,\dots, q)$

We partition the set of whole indices for the continuous variables in the previous subsection as $I = ((J, L), K) = ((O, L), Z) = (X, Z)$ and redefine the continuous variable itself as $\mathbf{x} = ((\mathbf{x}_J, \mathbf{x}_L), \mathbf{x}_K) \rightarrow ((\mathbf{x}_O, \mathbf{x}_L), \mathbf{z}) = (\mathbf{x}, \mathbf{z})$.
We also partition the set of whole indices for binary variables as $R=((S,U),T) = ((V, U), \emptyset)$, where $\emptyset$ is the empty set.
The index labels of the set of indices $O$ and $V$ may be understood as the initial letters of ``Observed'' and ``Visible'', respectively.
We further put the partitioned matrix and vector as $G_{RI} = (G_{RX}, G_{RZ}) = (0, G_{RZ})$ and $\mathbf{g}_{R I}^T = (\mathbf{g}_{RX}^T, \mathbf{g}_{RZ}^T) = (\bm{0}, \mathbf{g}_{RZ}^T)$ and redefining them as $G_{RZ} \rightarrow G$ and $\mathbf{g}_{RZ}^T \rightarrow \mathbf{g}_R^T$, respectively.
Then, we obtain
\begin{align}
    p(\mathbf{x}, \mathbf{z}, \mathbf{y}=\bm{1}_{R_1}) 
    =&
    \pi_{R_0}(\Sigma_{ZZ}) \, \mathcal{N}( \mathbf{x} \mid \bm{\mu}_X + \Sigma_{XZ} \Sigma_{ZZ}^{-1} (\mathbf{z} - \bm{\mu}_Z), \Sigma_{X|Z} ) 
    \mathcal{N}( \mathbf{z} \mid \bm{\mu}_Z - \Sigma_{ZZ} \mathbf{g}_{R_0}, \Sigma_{ZZ} ), \notag \\
    =&
    \pi_{R_1}(\Sigma_{ZZ}) \, \mathcal{N}( \mathbf{x} \mid \tilde{\bm{\mu}}_X + \Sigma_{XZ} \Sigma_{ZZ}^{-1} (\mathbf{z} - \tilde{\bm{\mu}}_Z), \Sigma_{X|Z} )
    \mathcal{N}( \mathbf{z} \mid \tilde{\bm{\mu}}_Z + \Sigma_{ZZ} G^T \mathbf{y}, \Sigma_{ZZ} ) , \\
    \pi_{R_0}(\Sigma_{ZZ})= &
    \frac{e^{ - \mathbf{b}_{R_0} + \frac{1}{2} \mathbf{g}_{R_0}^T \Sigma_{ZZ} \mathbf{g}_{R_0}} }{\sum_{R_0' \subseteq R} e^{ - \mathbf{b}_{R_0'} + \frac{1}{2} \mathbf{g}_{R_0'}^T \Sigma_{ZZ} \mathbf{g}_{R_0'}}}, \notag \\
    = \pi_{R_1}(\Sigma_{ZZ})= & 
    \frac{e^{ \bm{1}_{R_1}^T \tilde{\mathbf{b}} + \frac{1}{2} \bm{1}_{R_1}^T G \Sigma_{ZZ} G^T \bm{1}_{R_1}} }{\sum_{R_1' \subseteq R} e^{ \bm{1}_{R_1'}^T \tilde{\mathbf{b}} + \frac{1}{2} \bm{1}_{R_1'}^T G \Sigma_{ZZ} G^T \bm{1}_{R_1'}}},
\end{align}
\begin{align}
    p(\mathbf{z}) =& 
    \sum_{R_0  \subseteq R} \pi_{R_0}(\Sigma_{ZZ}) \, \mathcal{N}( \mathbf{z} \mid \bm{\mu}_Z - \Sigma_{ZZ} \mathbf{g}_{R_0}, \Sigma_{ZZ} ), \notag \\
    =& 
    \sum_{R_1  \subseteq R} \pi_{R_1}(\Sigma_{ZZ}) \, \mathcal{N}( \mathbf{z} \mid \tilde{\bm{\mu}}_Z + \Sigma_{ZZ} G^T \bm{1}_{R_1}, \Sigma_{ZZ} ), \\
    p(\mathbf{x}_O, \mathbf{y}_V|\mathbf{z}) =&
    \frac{\sum_{U_0 \subseteq R \setminus V} e^{ - \mathbf{b}_{R_0} -\mathbf{g}_{R_0}^T (\mathbf{z} -\bm{\mu}_Z)}}{\sum_{R_0' \subseteq R}  e^{ - \mathbf{b}_{R_0'} -\mathbf{g}_{R_0'}^T (\mathbf{z} - \bm{\mu}_Z)}}
     \,
    \mathcal{N}( \mathbf{x}_O \mid \bm{\mu}_O + \Sigma_{OZ} \Sigma_{ZZ}^{-1} (\mathbf{z}-\bm{\mu}_Z), \Sigma_{O|Z} ), \notag \\
    =& 
    \frac{\sum_{U_1 \subseteq R \setminus V} e^{ \bm{1}_{R_1}^T (\tilde{\mathbf{b}} + G (\mathbf{z} - \tilde{\bm{\mu}}_Z))}}{\sum_{R_1' \subseteq R} e^{\bm{1}_{R_1'}^T (\tilde{\mathbf{b}} + G (\mathbf{z} - \tilde{\bm{\mu}}_Z)) }}
     \,
    \mathcal{N}( \mathbf{x}_O \mid \tilde{\bm{\mu}}_O + \Sigma_{OZ} \Sigma_{ZZ}^{-1} (\mathbf{z}-\tilde{\bm{\mu}}_Z), \Sigma_{O|Z} ).
\end{align}

In a similar way, we partition the set of whole indices for continuous variables in the previous subsection as $I=((J, L), K) = ((Z, L), O)$ and redefine the continuous variable itself as $\mathbf{x} = ((\mathbf{x}_J, \mathbf{x}_L), \mathbf{x}_K) \rightarrow ((\mathbf{z},\mathbf{x}_L),\mathbf{x}_O)$.
We also partition the set of whole indices for binary variables as $R = ((S,U),T) = ((\emptyset, U), V)$.
Putting the partitioned matrix and vector as $G_{RI} = (G_{RZ}, G_{R(L+O)}) = (G_{RZ}, 0)$ and $\mathbf{g}_{R I}^T = (\mathbf{g}_{RZ}^T, \mathbf{g}_{R (L+O)}^T) = (\mathbf{g}_{RZ}^T, \bm{0})$ and redefining them as $G_{RZ} \rightarrow G$ and $\mathbf{g}_{RZ}^T \rightarrow \mathbf{g}_R^T$, we obtain
\begin{align}
    p(\mathbf{x}_O, \mathbf{y}_V) =& 
    \sum_{U_0 \subseteq R \setminus V} \pi_{R_0}(\Sigma_{ZZ}) \, 
    \mathcal{N}( \mathbf{x}_O \mid \bm{\mu}_O - \Sigma_{OZ} \mathbf{g}_{R_0}, \Sigma_{OO} ), \notag \\
    =& 
    \sum_{U_1 \subseteq R \setminus V} \pi_{R_1}(\Sigma_{ZZ}) \,
    \mathcal{N}( \mathbf{x}_O \mid \tilde{\bm{\mu}}_O + \Sigma_{OZ} G^T \bm{1}_{R_1}, \Sigma_{OO} ),
\end{align}
\begin{align}
    p(\mathbf{z}|\mathbf{x}_O, \mathbf{y}_V) =& \frac{\sum_{U_0 \subseteq R \setminus V}  e^{- \mathbf{b}_{R_0} + \frac{1}{2}\mathbf{g}_{R_0}^T \Sigma_{Z|O} \mathbf{g}_{R_0}} e^{-\mathbf{g}_{U_0}^T \Sigma_{ZO} \Sigma_{OO}^{-1} (\mathbf{x}_O - \bm{\mu}_O)}}{\sum_{U_0' \subseteq R \setminus V} e^{ - \mathbf{b}_{R_0'} + \frac{1}{2} \mathbf{g}_{R_0'}^T \Sigma_{Z|O} \mathbf{g}_{R_0'}} e^{-\mathbf{g}_{U_0'}^T \Sigma_{ZO} \Sigma_{OO}^{-1} (\mathbf{x}_O-\bm{\mu}_O)}} \notag \\
    & \mathcal{N}\bigl( \mathbf{z} \mid \bm{\mu}_Z + \Sigma_{ZO}\Sigma_{OO}^{-1}(\mathbf{x}_O - \bm{\mu}_O) - \Sigma_{Z|O} \mathbf{g}_{R_0}, \Sigma_{Z|O} \bigr) , \notag \\
    =& 
    \frac{\sum_{U_1 \subseteq R \setminus V} e^{\bm{1}_{U_1}^T \tilde{\mathbf{b}} + \frac{1}{2} (\bm{1}_{U_1}^T + \bm{1}_{V_1})^T G \Sigma_{Z|O} G^T (\bm{1}_{U_1} + \bm{1}_{V_1}) + \bm{1}_{U_1}^T G \Sigma_{ZO} \Sigma_{OO}^{-1} (\mathbf{x}_O - \tilde{\bm{\mu}}_O)}}{\sum_{U_1' \subseteq R \setminus V} e^{\bm{1}_{U_1'}^T \tilde{\mathbf{b}} + \frac{1}{2} (\bm{1}_{U_1'}^T + \bm{1}_{V_1})^T G \Sigma_{Z|O} G^T (\bm{1}_{U_1'} + \bm{1}_{V_1}) + \bm{1}_{U_1'}^T G \Sigma_{ZO} \Sigma_{OO}^{-1} (\mathbf{x}_O - \tilde{\bm{\mu}}_O)}} \notag \\
    &\mathcal{N}\bigl( \mathbf{z} \mid \tilde{\bm{\mu}}_Z + \Sigma_{ZO}\Sigma_{OO}^{-1}(\mathbf{x}_O - \tilde{\bm{\mu}}_O) + \Sigma_{Z|O} G^T (\bm{1}_{U_1} + \bm{1}_{V_1}), \Sigma_{Z|O} \bigr).
\end{align}

If the observed variables have no missing values, the above expressions can be expressed more concisely:
\begin{align}
    p(\mathbf{x}, \mathbf{z}, \mathbf{y}) =& \pi_{R_1}(\Sigma_{ZZ}) \, \mathcal{N}\bigl( (\mathbf{x}^T, \mathbf{z}^T)^T \mid \tilde{\bm{\mu}}_{(X+Z)} + \Sigma_{(X+Z) Z} G \bm{1}_{R_1} \bigr), \notag \\
    =& 
    \pi_{R_1}(\Sigma_{ZZ}) \, 
    \mathcal{N}( \mathbf{x} \mid \tilde{\bm{\mu}}_X + \Sigma_{XZ} \Sigma_{ZZ}^{-1}(\mathbf{z}- \tilde{\bm{\mu}}_Z), \Sigma_{X|Z} ) \, 
    \mathcal{N}( \mathbf{z} \mid \tilde{\bm{\mu}}_Z + \Sigma_{ZZ} G^T \mathbf{y}, \Sigma_{ZZ} ), \\
    p(\mathbf{z}) =& 
    \sum_{R_1  \subseteq R} \pi_{R_1}(\Sigma_{ZZ}) \,
    \mathcal{N}( \mathbf{z} \mid \tilde{\bm{\mu}}_Z + \Sigma_{ZZ} G^T \bm{1}_{R_1}, \Sigma_{ZZ} ), \\
    p(\mathbf{x}, \mathbf{y}|\mathbf{z}) =& 
    \frac{e^{ \mathbf{y}^T (\tilde{\mathbf{b}} + G (\mathbf{z}-\tilde{\bm{\mu}}_Z))}}{\prod_{j=1}^q \bigl( 1 + e^{\tilde{b}_j + \mathbf{g}_j^T (\mathbf{z} - \tilde{\bm{\mu}}_Z)} \bigr) } \, \mathcal{N}\bigl(\mathbf{x} \mid \tilde{\bm{\mu}}_X + \Sigma_{XZ} \Sigma_{ZZ}^{-1} (\mathbf{z}-\tilde{\bm{\mu}}_Z), \Sigma_{X|Z} \bigr), \\
    p(\mathbf{x}, \mathbf{y}) =& \pi_{R_1}(\Sigma_{ZZ}) \, \mathcal{N}(\mathbf{x} \mid \tilde{\bm{\mu}}_X + \Sigma_{XZ} G^T \mathbf{y}, \Sigma_{XX} ), \\
    p(\mathbf{z}|\mathbf{x}, \mathbf{y}) =& \mathcal{N}\bigl(\mathbf{z} \mid \tilde{\bm{\mu}}_Z + \Sigma_{ZX}\Sigma_{XX}^{-1}(\mathbf{x} - \tilde{\bm{\mu}}_X) + \Sigma_{Z|X} G^T \mathbf{y}, \Sigma_{Z|X} \bigr), \\
    \pi_{R_1}(\Sigma_{ZZ}) =&
    \frac{e^{ \bm{1}_{R_1}^T \tilde{\mathbf{b}} + \frac{1}{2} \bm{1}_{R_1}^T G \Sigma_{ZZ} G^T \bm{1}_{R_1}} }{\sum_{R_1' \subseteq R} e^{ \bm{1}_{R_1'}^T \tilde{\mathbf{b}} + \frac{1}{2} \bm{1}_{R_1'}^T G \Sigma_{ZZ} G^T \bm{1}_{R_1'}}}.
\end{align}
To derive the expressions in the main text of the paper, we parameterize the covariance matrix for continuous variables by a block partitioned matrix as follows:
\begin{align}
    \Sigma^{-1} =&
    \begin{bmatrix} \Sigma_{XX} & \Sigma_{XZ} \\ \Sigma_{ZX} & \Sigma_{ZZ} \end{bmatrix}^{-1} 
    \equiv \begin{bmatrix} \Sigma_{x} & \Sigma_{xz} \\ \Sigma_{zx} & \Sigma_{z} \end{bmatrix}^{-1}, \notag \\
    = & \begin{bmatrix} \Psi + W \Sigma_z W^T & W \Sigma_z \\ \Sigma_z W^T & \Sigma_z \end{bmatrix}^{-1}
    = \left(
    \begin{bmatrix}
        \Psi^{1/2} & W \Sigma_z^{1/2} \\ O & \Sigma_z^{1/2} 
    \end{bmatrix}
    \begin{bmatrix}
        \Psi^{1/2} & O \\ \Sigma_z^{1/2} W^T & \Sigma_z^{1/2}
    \end{bmatrix}
    \right)^{-1}, \\
    =& \begin{bmatrix} \Psi^{-1} & - \Psi^{-1} W \\ - W^T \Psi^{-1} & \Sigma_z^{-1} + W^T \Psi^{-1} W \end{bmatrix},
\end{align}
where $O$ is a matrix with all elements zero and $\Psi$ is a diagonal matrix with non-negative diagonal elements.
That is, $\Sigma_{x} = \Psi + W \Sigma_z W^T$, $\Sigma_{xz} = W \Sigma_z$, $\Sigma_{x|z} = \Psi$, $\Sigma_{z|x} = [\Sigma_z^{-1} + W^T \Psi^{-1} W]^{-1}$, where the notation $\Sigma_{x|z} \equiv \Sigma_x - \Sigma_{xz} \Sigma_z^{-1} \Sigma_{zx}$ means the Schur complement.
If we assume that $\Sigma_z$ is a diagonal matrix with non-negative diagonal elements, then the above expression can be interpreted as a Cholesky decomposition of $\Sigma$, which means that $\Sigma_x$ is by itself a positive semi-definite matrix.
Redefining the parameters as $\tilde{\mathbf{b}} \rightarrow \mathbf{b}$, $\tilde{\bm{\mu}} \rightarrow \bm{\mu}$, and $\bm{\mu}=(\bm{\mu}_X, \bm{\mu}_Z) \rightarrow (\bm{\mu}_x, \bm{\mu}_z)$, we obtain the expressions in the main text of the paper, Eqs.~(\ref{eq:fa_conditional}, \ref{eq:fa_prior}, \ref{eq:fa_induced}, \ref{eq:fa_posterior}, \ref{eq:fa_joint}).

\section*{References}

\bibliography{continuous_binary}

\end{document}